\documentclass[twocolumn,prd,aps,tightenlines,floats,floatfix,preprintnumbers,nofootinbib,eqsecnum]{revtex4}%%

\def\sp{\kern +3pt}
\def\sm{\kern -3pt}
\def\spQ{\kern +6pt}

\def\bea{\begin{eqnarray}}
\def\eea{\end{eqnarray}}

\def\sfrac#1#2{{\textstyle \frac{#1}{#2}}}

\def\be{\begin{equation}}
\def\ee{\end{equation}}
\def\ba{\begin{eqnarray}}
\def\ea{\end{eqnarray}}

\usepackage{graphics}
\usepackage{graphicx}
\usepackage{epsf}
\usepackage{amsmath}
\usepackage{amssymb}
\usepackage{bbold}
\usepackage{xcolor}

\begin{document}

\phantom{0}
\vspace{-0.2in}
\hspace{5.5in}

% include preprint number option
\preprint{{\bf  LFTC-19-11/49}}
%\preprint{{\bf  Version arXiv}}
%\preprint{{\bf  Version 13}}

\vspace{-1in}%\parbox{1.5in}{ \vspace{-9.6in}}  % moves the preprint box down

\title
{\bf Low-$Q^2$ empirical parametrizations of the $N^\ast$ 
helicity amplitudes}
\author{G.~Ramalho}
\vspace{-0.1in}

\affiliation{Laborat\'orio de 
F\'{i}sica Te\'orica e Computacional -- LFTC,
Universidade Cruzeiro do Sul/Universidade Cidade de  S\~ao Paulo,  \\
01506-000,   S\~ao Paulo, SP, Brazil}

\vspace{0.2in}
\date{\today}

\phantom{0}

\begin{abstract}
The data associated with the electromagnetic excitations 
of the nucleon ($\gamma^\ast N \to N^\ast$) are usually 
parametrized by helicity amplitudes at the resonance $N^\ast$ rest frame.
The properties of the $\gamma^\ast N \to N^\ast$ transition current
at low $Q^2$  can be, however, better understood when expressed in terms 
of structure form factors,
particularly near the pseudothreshold, when the magnitude of the photon 
three-momentum vanishes ($|{\bf q}| =0$). 
At the pseudothreshold
the invariant four-momentum square became $q^2= (M_R-M_N)^2$,
well in the timelike region $Q^2 =- q^2 < 0$
[$M_N$ and $M_R$ are the mass of the nucleon and of 
the resonance, respectively].
In the helicity amplitude representation,
the amplitudes have well-defined dependences on  $|{\bf q}|$,
near the pseudothreshold, and there are correlations between different amplitudes.
Those constraints are often ignored in the empirical 
parametrizations of the  helicity amplitudes.
In the present work, we show that the structure of 
the transition current near the pseudothreshold has an impact 
on the parametrizations of the data.
We present a method which modifies analytic parametrizations of the data 
at low $Q^2$, in order to take into account 
the constraints of the transition amplitudes near the pseudothreshold.
The model dependence of the parametrizations 
on the low-$Q^2$ data is studied in detail.
\end{abstract}

%\phantom{0}
%\vspace{7.0in}
%\vspace{-6in}
\vspace*{0.9in}  % sets how far the title is below the preprint box
\maketitle

\section{Introduction}
\label{secIntro}

In the past two decades, there was significant 
experimental progress in the study of the electromagnetic 
structure of the nucleon ($N$) and the nucleon excited states 
($N^\ast$)~\cite{NSTAR,Aznauryan12b,Puckett17,Drechsel2007}.
From the theoretical point of view 
there was some progress in the description 
of the electromagnetic excitations $N^\ast$ using 
QCD sum rules~\cite{Lenz09,Wang09,Anikin15},
approaches based on the Dyson-Schwinger equations~\cite{Eichmann16,Segovia16},
AdS/QCD~\cite{Teramond12,Brodsky15,Bayona12,Ramalho17,Gutsche} and 
different classes of quark models~\cite{Obukhovsky,AznauryanII,Santopinto,NstarP1,SQTM}.
At low $Q^2$, the dynamical coupled-channel models which 
take into account the meson cloud dressing of the baryons are also 
successful~\cite{Kamalov99,Drechsel2007,Burkert04,JDiaz07}.

The electromagnetic structure of the nucleon excitations
can be probed through the scattering of electrons on nucleons ($e N \to e^\prime N^\ast$).
The measured  cross sections include the 
information about the $\gamma^\ast N \to N^\ast$ transitions, which can be expressed 
in terms of different structure functions 
depending on the photon polarization and on the 
four-momentum transfer squared, $q^2$.
The most common representation of those structure functions 
is the helicity amplitude representation, 
where all the  $\gamma^\ast N \to N^\ast$ transitions 
are parametrized in terms of the three different 
polarizations of the photon, 
including  two transverse amplitudes $A_{1/2}$ and $A_{3/2}$ and 
one longitudinal amplitude $S_{1/2}$,
depending on the angular momentum $J$ of the nucleon resonance $N^\ast$.
Those amplitudes are usually presented in the 
rest frame of the resonance ($N^\ast$)~\cite{Aznauryan12b,Bjorken66,Jones73,Devenish76,Drechsel2007}.

An alternative representation of those structure functions
is the transition form factor representation, generally 
derived from the structure of the $\gamma^\ast N \to N^\ast$ 
transition current~\cite{Aznauryan12b,Bjorken66,Jones73,Devenish76,Compton}.
Examples of the form factor representation 
are the Dirac and Pauli form factors,
the electric and magnetic (Sachs) form factors of the nucleon~\cite{Puckett17}.
The nucleon resonances can also be described 
by the Dirac- and Pauli-like form factors for the $J=\frac{1}{2}^\pm$ resonances
or multipole form factors, such as the electric, magnetic 
and Coulomb form factors for the $J=\frac{3}{2}^\pm$ resonances, 
with positive or negative parity 
($P=\pm$)~\cite{Jones73,Devenish76,Bjorken66,Compton,Siegert1,Siegert2}.
The form factor representation has some advantages in
the interpretation of the structure of the $J^\pm$  states,
since it emphasizes the symmetries associated with 
the nucleon resonances~\cite{Compton}.

Empirical parametrizations of the helicity amplitudes 
or transition form factors are important 
to compare with theoretical models and for 
calculations which require 
accurate descriptions of the $N^\ast$ data,
in different ranges of $Q^2$~\cite{Compton,Blin19a}.

In general, the $J^\pm$ helicity amplitudes or the 
transition form factors are independent functions of $Q^2=-q^2$,
except in some particular kinematic limits.
An important limit is the pseudothreshold limit,
where the magnitude of photon three-momentum $|{\bf q}|$ vanishes,
and the nucleon and the nucleon resonance are both at rest.
In this limit, the invariant $Q^2$ has the value 
$Q^2= -(M_R -M_N)^2$, where $M_N$  and $M_R$ 
are the mass of the nucleon and the nucleon resonance $N^\ast$, respectively.
Throughout this work, we also use
$R$ to label $N^\ast$ and the properties of $N^\ast$.

The transition current associated with states $J=\frac{1}{2}, \frac{3}{2}$ 
can be decomposed into two $\left(J= \frac{1}{2}\right)$ or three 
$\left(J= \frac{3}{2}\right)$ gauge-invariant terms associated 
with the properties  of the transition. 
Each of those terms defines elementary form factors $G_i$ 
independent of each other and free 
of kinematical singularities~\cite{Jones73,Devenish76,Bjorken66}.
Helicity amplitudes in a given frame and the 
transition form factors (Dirac-Pauli or multipole)
can be expressed 
as linear combinations 
of the elementary form factors $G_i$~\cite{Aznauryan12b,Devenish76,Bjorken66}.
The helicity amplitudes 
and transition form factors are not uncorrelated
as the elementary form factors, 
because of the kinematical factors 
included in the transformation between the elementary form 
factors and the alternative structure functions.
The discussion of those transformations 
and their consequences in the pseudothreshold limit 
can be found in Refs.~\cite{Jones73,Devenish76,Bjorken66}.

When we parametrize the data in term of the elementary 
form factors $G_i$, we do not need to worry 
about the constraints at the pseudothreshold, 
since those are automatically ensured
and the structure functions are free of 
kinematical singularities~\cite{Devenish76}.
Consistent parametrizations of the helicity amplitudes based 
on appropriated elementary form factors for 
the $J= \frac{1}{2}, \frac{3}{2}$ resonances 
are presented in Ref.~\cite{Compton}.

When we parametrize the transition form factors 
or the helicity amplitudes at the resonance rest frame directly,
however,  we need to take into account the pseudothreshold constraints.
Those structure functions are constrained by   
some specific dependence on $|{\bf q}|$~\cite{Drechsel92,AmaldiBook,Tiator16} 
and by correlations between different helicity amplitudes
and, equivalently, by correlations between 
transition form factors~\cite{Jones73,Devenish76,Bjorken66}.
Those constraints are the consequence of the gauge-invariant structure 
of the transition current and the kinematics associated with the 
$N^\ast$ rest frame~\cite{Devenish76,Bjorken66}.

The dependence of the transverse amplitudes ($A_{1/2}$ and $A_{3/2}$)
and longitudinal amplitudes  in the magnitude of 
the photon three-momentum $|{\bf q}|$ 
for the  cases  $J= \frac{1}{2}^\pm$, $\frac{3}{2}^\pm$ 
is summarized in Table~\ref{tableST1}.
Additional discussion about those relations 
can be found in 
Refs.~\cite{Bjorken66,Devenish76,Jones73,Siegert1,Siegert2,Compton,Drechsel92}.
The content of the table shows that the constraints on the helicity amplitudes 
cannot be ignored if the pseudothreshold of 
the $\gamma^\ast N \to N^\ast$ transition is close to the photon point $Q^2=0$.
This observation demonstrates the need of taking into account 
the pseudothreshold constraints in 
the empirical parametrizations of the helicity amplitudes, particularly 
for resonances with masses close to the nucleon.

\begin{table}[t]
\begin{center}
\begin{tabular}{c | c   c   c }
\hline
\hline 
\vspace{.1cm}
$\frac{1}{2}^+$ & $A_{1/2} \propto|{\bf q}|$, 
&  $S_{1/2} \propto|{\bf q}|^2 $  &  \\[.1cm]
$\frac{1}{2}^-$ & $A_{1/2} \propto 1 $, &  $S_{1/2} \propto|{\bf q}| $  &   $S_{1/2} \propto A_{1/2} |{\bf q}|$\\[.4cm]
%\hline
$\frac{3}{2}^+$ & $A_{1/2} \propto|{\bf q}|$, 
&  $S_{1/2} \propto|{\bf q}|^2 $  & \spQ \spQ $  S_{1/2}  \propto (A_{1/2} - \sfrac{1}{\sqrt{3}} A_{3/2})|{\bf q}| $ \\[.1cm]
                & $A_{3/2} \propto|{\bf q}|$, &                         &    
%$(A_{1/2} + \sqrt{3} A_{3/2}) \propto |{\bf q}|$                  
\\[.1cm]
$\frac{3}{2}^-$ & $A_{1/2} \propto 1$, &  $S_{1/2} \propto|{\bf q}| $  
& $    S_{1/2}   \propto (A_{1/2} + \sqrt{3} A_{3/2})|{\bf q}| $  \\[.1cm]
                & $A_{3/2} \propto 1$, &                         &   
$(A_{1/2} - \sfrac{1}{\sqrt{3}} A_{3/2}) \propto |{\bf q}|^2$ \\[.1cm]          
\hline
\hline
\end{tabular}
\end{center}
\caption{\footnotesize
Constraints at the pseudothreshold.}
\label{tableST1}
\end{table}

The present work is motivated by the necessity 
of taking into account the constraints from Table~\ref{tableST1}
in the empirical parametrization of the data.
Since the empirical parametrizations of the data ignore, 
in general, the specific dependence on $|{\bf q}|$,
in the present work we checked if it is possible 
to modify those parametrizations 
below a certain value of $Q^2$, labeled as $Q_P^2$, 
in order to obtain a consistent extension to the pseudothreshold, 
without spoiling the description of the available data.
We derive then analytic extensions of 
generic parametrizations of the data based on the continuity 
of the amplitudes and on the continuity of the first derivatives 
of those amplitudes in the transition point $Q_P^2$.
With this procedure, we generate smooth extensions of available 
parametrizations of the data, and study the consistence 
of the results with the data and with the pseudothreshold constraints.
The solutions obtained can also be used to test the 
sensibility of the solutions to possible variations 
of the data at low $Q^2$.
This study is particularly useful, since there 
is generally a gap in the data between $Q^2=0$ and $Q^2=0.3$ GeV$^2$.

The formalism proposed to extend analytically the amplitudes 
to the timelike region is 
general and can be applied to any  
set of parametrizations of amplitudes,
provided that those are continuous and that 
their first derivatives are also continuous  
in the spacelike region ($Q^2 \ge 0$).
To exemplify our formalism, we consider a particular set 
of empirical parametrizations of the $\gamma^\ast N \to N^\ast$  
helicity amplitudes,
associated with the $N^\ast$ states:
$N(1440)$, $N(1520)$, $N(1535)$,
$N(1650)$, $N(1710)$, $N(1720)$,  $\Delta(1232)$, $\Delta(1620)$ and $\Delta(1700)$.
We look, in particular, to the Jefferson Lab parametrizations  
from Ref.~\cite{Jlab-fits}, 
which have been used by several authors.
Those parametrizations are based on simple expressions 
(rational functions of $\sqrt{Q^2}$), 
are, in general, valid in the range $Q^2=0.5$--5 GeV$^2$, 
and provide a fair description of the available data~\cite{MokeevDatabase}, 
in particular of the CLAS/JLab data~\cite{CLAS09,CLAS12,Mokeev16,Blin19a} 
at intermediate and large $Q^2$.

This article is organized as follows:
In the next section, we discuss in more detail 
the implications of the pseudothreshold conditions 
on the helicity amplitudes.
In Sec.~\ref{secEmpirical}, we review the definition of the 
helicity amplitudes in the $N^\ast$ rest frame, 
and discuss which information is necessary to 
derive an analytic continuation of a given amplitude.
The method used to extend the empirical parametrization to the timelike
region, including the pseudothreshold limit,
is presented in Sec.~\ref{secAnalytic}.
The numerical results for all $N^\ast$ states considered
are presented and analyzed in Sec.~\ref{secResults}.
In Sec.~\ref{secConclusions} we present 
our outlook and conclusions.

\section{Constraints on the Helicity amplitudes}
\label{secConstraints}

The most popular consequence of the pseudothreshold constraints 
is the relation between the electric amplitude $E$ 
(combination of longitudinal amplitudes) and the scalar amplitude $S_{1/2}$
which can, in general, be expressed in the form
\ba
\lambda_R \, S_{1/2}  = E  \, |{\bf q}|,
\label{eqST}
\ea
where the factor $\lambda_R \propto (M_R -M_N)$ 
depends on the masses of the nucleon and the resonance.
The relation (\ref{eqST}) is known as Siegert's theorem or 
the long-wavelength theorem, 
since it is valid in the limit $|{\bf q}| \to 0$~\cite{Drechsel92,Buchmann98,Tiator16}.
There are, however, other constraints, associated with 
the specific dependence of the amplitudes on $|{\bf q}|$, displayed in Table~\ref{tableST1}.

One can illustrate the importance of 
including the correct $|{\bf q}|$ dependence
on the amplitudes, looking for the simplest case, 
the $\frac{1}{2}^+$ helicity amplitudes.
Those amplitudes can be written as~\cite{Aznauryan12b,Roper} 
\ba
& &
\hspace{-.5cm}
A_{1/2} = {\cal R} \frac{|{\bf q}|}{\sqrt{1 + \tau}} 
(\tau G_1 + G_2) \nonumber \\
& &
\hspace{-.5cm}
S_{1/2} = \frac{1}{\sqrt{2}(M_R + M_N)} {\cal R}
\frac{|{\bf q}|^2}{\sqrt{1 +\tau}} ( G_1 -  G_2),
\label{eqRoper}
\ea
where ${\cal R}$ is a constant\footnote{${\cal R}$ can be represented 
as 
$$
{\cal R} = \frac{e}{\sqrt{M_N M_R K}} \frac{2 M_R}{M_N + M_R},
$$
where $e$ is the elementary electric charge and $K= \frac{M_R^2- M_N^2}{2 M_R}$.},
$\tau = \frac{Q^2}{(M_R + M_N)^2}$,
$\tau G_1$ represents the Dirac form factor
and $G_2$ represents the Pauli form factor.
Based on the previous representation, we conclude 
that if $G_1$ and $G_2$ are regular functions, with no zeros 
and singularities at the pseudothreshold,
one has automatically $A_{1/2} \propto |{\bf q}|$ 
and $S_{1/2} \propto |{\bf q}|^2$.
Most of the empirical parametrizations of the data based on parametrizations of 
the amplitudes $A_{1/2}$ and $S_{1/2}$ by regular functions, 
ignore this specific dependence on $|{\bf q}|$.

One concludes, then, that  if the form factors ($G_1$ and $G_2$)
are not parametrized directly, we need to enforce 
the dependence of the amplitudes on $|{\bf q}|$, in order to have a consistent 
parametrization of the data, based on the properties of the transition currents.
Empirical parametrizations of the data which ignore the correct  
$|{\bf q}|$ dependence are inconsistent and 
provide erroneous descriptions of 
the helicity amplitudes at low $Q^2$.

There are in the literature several works which explore 
the relations between helicity amplitudes and transition form factors 
and try to identify in the data, signatures of the pseudothreshold constraints.
The $\gamma^\ast N \to \Delta(1232)$ have been discussed in some detail 
by several authors~\cite{Buchmann97a,Buchmann98,Buchmann04,Tiator16,Drechsel2007,Siegert1,Siegert2,DeltaQuad,RSM-Siegert,GlobalP}.
The  $\gamma^\ast N \to N(1535)$ and  $\gamma^\ast N \to N(1520)$ 
are also discussed in Refs.~\cite{Tiator16,Siegert1,Siegert2}.
As for the states $\frac{1}{2}^+$, as the Roper,
one can conclude from Table.~\ref{tableST1}, 
that there is no relation associated with Siegert's theorem.
This happens because for the $\frac{1}{2}^+$ states
there is no electric amplitude~\cite{Drechsel92,Drechsel2007}.
Nevertheless, there are constraints for the amplitudes $A_{1/2}$ and $S_{1/2}$,
which cannot be ignored.

Instead of deriving alternative parametrizations of the helicity amplitudes consistent 
with the pseudothreshold constraints, in the present work, 
we propose a method to modify available parametrizations of the data
in order to satisfy those constraints.
The consistence of the modified parametrizations 
with the low-$Q^2$ data can be checked afterwards.

\section{Empirical parametrizations of Helicity amplitudes}
\label{secEmpirical}

The experimental data associated with the 
$\gamma^\ast N \to N^\ast$ transitions are usually represented 
in term of the helicity amplitudes in the $N^\ast$ rest frame.
Those amplitudes can be calculated from the 
transition current $J^\mu$,
in units of the elementary charge $e$,
using~\cite{Aznauryan12b}:
\ba
A_{3/2}&=&
\sqrt{\frac{2\pi \alpha}{K}}
\left< R, S_z'=+\sfrac{3}{2} \right|
\epsilon_+ \cdot J \left| N,S_z=+\sfrac{1}{2} \right> 
\nonumber  \\
& &
\label{eqA32} \\
A_{1/2}
&= &\sqrt{\frac{2\pi \alpha}{K}}
\left< R, S_z'=+\sfrac{1}{2} \right|
\epsilon_+ \cdot J \left| N,S_z=-\sfrac{1}{2} \right> 
\nonumber \\
\label{eqA12} \\
S_{1/2}&=&
\sqrt{\frac{2\pi \alpha}{K}}
\left< R, S_z'=+\sfrac{1}{2} \right|
\epsilon_0 \cdot J \left| N,S_z=+\sfrac{1}{2} \right>\frac{|{\bf q}|}{Q},
\nonumber \\
& &
\label{eqS12}
\ea
where $S_z'$ ($S_z$) is the final (initial) 
spin projection,
${\bf q}$ is the photon three-momentum in the rest frame of $R$,
$Q=\sqrt{Q^2}$ and   $\epsilon_\lambda^\mu$ ($\lambda=0,\pm 1$) are
the photon polarization vectors.
In the previous equations
 $\alpha \simeq 1/137$ 
is the fine-structure constant  and 
$K= \frac{M_R^2-M_N^2}{2M_R}$
is the magnitude of the photon momentum when $Q^2=0$.
In the rest frame of $R$ the magnitude of the nucleon three-momentum is
 $|{\bf q}|$, and reads
\be
|{\bf q}|= \frac{\sqrt{Q_+^2Q_-^2}}{2M_R},
\label{eqq2}
\ee
where $Q_\pm^2= (M_R \pm M_N)^2 + Q^2$.

Depending on the spin $J$ of the resonance one can have 
two ($J=\frac{1}{2}$) or three ($J=\frac{3}{2}$) independent amplitudes.
There are in the literature several kinds of parametrizations of the data.
The MAID (Mainz Unitary Isobar Model) 
parametrizations are characterized by 
the combination of polynomials and exponentials~\cite{Drechsel2007,Siegert1,Siegert2},
and other parametrizations 
are based on rational functions~\cite{Compton,Aznauryan12b,Siegert2}.
In principle, all those parametrizations 
are equivalent in a certain range of $Q^2$, 
provided that $Q^2 \ge 0 $ and that we are not too 
close to the pseudothreshold.

The important for the following discussion 
is that the parametrization of a generic amplitude $A$
corresponds to a regular function of $Q^2$
(no zeros at $Q^2=0$ and no singularities),
that the function is continuous and that 
the first derivatives exist and are also continuous.

We assume then that the parametrizations under discussion
take known analytic forms, and describe well the data 
above a given threshold $Q_P^2 > 0$.
In those cases  we can check if we can derive 
analytic continuations to the timelike region 
consistent with the pseudothreshold constraints and 
with smooth transitions between the pseudothreshold  
$Q^2= -(M_R-M_N)^2 $ and the point $Q^2=  Q_P^2$.

By varying the value of $Q_P^2$, we infer the sensibility 
of the fits to the pseudothreshold conditions.
Since the empirical parametrizations of the data 
can be in some cases very sensitive to the low-$Q^2$ data, 
and also because there is in most resonances a gap between
$Q^2=0$ and $Q^2=0.3$ GeV$^2$, we consider 
three different values for  $Q_P^2$: $Q_P^2=0.1$ 0.3 and 0.5 GeV$^2$.
We avoid intentionally the use of the $Q^2=0$,
in order to derive an analytic continuation independent of the $Q^2=0$ data.

The analytic continuation is derived demanding 
a smooth transition between the region  
where the original parametrization is assumed to be valid,
the $Q^2 \ge Q_P^2$ region, and the region between 
the pseudothreshold, $Q^2= -(M_R-M_N)^2$, 
and the point $Q^2= Q_P^2$.
Below  $Q^2= Q_P^2$ the parametrizations 
are consistent with the expected shape near the pseudothreshold,
characterized by the expressions from Table~\ref{tableST1}.
The smooth transition between the two regions is 
obtained by imposing the following conditions:
\begin{itemize}
\item
The amplitude $A$ is continuous at $Q^2=Q_P^2$;
\item
the first derivative is continuous at $Q^2=Q_P^2$; and 
\item 
the second derivative is continuous at $Q^2=Q_P^2$.
\end{itemize}
In some cases, we demand also 
the continuity of the third derivative at $Q^2=Q_P^2$.

To derive the analytic continuation, we consider the expansion
\ba
A(Q^2) &=&  A^{(0)} + A^{(1)} (Q^2 -Q_P^2)  + \nonumber \\ 
      & & \frac{A^{(2)}}{2!} (Q^2  -Q_P^2)^2  + \frac{A^{(3)}}{3!}  (Q^2  -Q_P^2)^3 + ...., 
\nonumber \\
\label{eqAmp1}
\ea
where the coefficients $A^{(k)}$ ($k=1,2,3$) represent the derivatives 
$A^{(k)} =  \frac{d^k A}{d Q^{2k}} (Q_P^2)$ 
and $A^{(0)} = A(Q_P^2)$.
In the following, we refer to $A^{(k)}$ as the moments 
of the expansion of $A$ in $Q^2$.

The analytic continuation for the $ - (M_R -M_N)^2 <Q^2  < Q_P^2$ region 
is discussed in the next section.

\begin{table*}[t]
\begin{center}
\begin{tabular}{c c c c }
\hline
\hline 
%\vspace{.1cm}
$\frac{1}{2}^+$ \spQ
& 
$A_{1/2} = a_0 \tilde q  + a_1  \tilde q^3 + a_2  \tilde q^5 + a_3  \tilde q^7$ 
\sp
&  &
$S_{1/2} = c_0 \tilde q^2 + c_1  \tilde q^4 + c_2  \tilde q^6 + c_3 \tilde q^8$ 
\\[.1cm]
$\frac{1}{2}^-$  \spQ
& 
$A_{1/2} = a_0   + a_1  \tilde q^2 + a_2  \tilde q^4 + a_3  \tilde q^6$ \sp
& &
$S_{1/2} = {\bf c_0} \tilde q + c_1  \tilde q^3 + c_2  \tilde q^5 + c_3 \tilde q^5$ \\[.3cm]
%& & & \\[.1cm]
$\frac{3}{2}^+$ \spQ
& 
$A_{1/2} = a_0 \tilde q  + a_1  \tilde q^3 + a_2  \tilde q^5 + a_3  \tilde q^7$ 
\sp 
&
$A_{3/2} = b_0 \tilde q  + b_1  \tilde q^3 + b_2  \tilde q^5 + b_3  \tilde q^7$ 
\sp
& 
$S_{1/2} = {\bf c_0} \tilde q^2 + c_1  \tilde q^4 + c_2  \tilde q^6 + c_3 \tilde q^8$ \\[.1cm]
$\frac{3}{2}^-$ \spQ
& 
$A_{1/2} = a_0   + a_1  \tilde q^2 + a_2  \tilde q^4 + a_3  \tilde q^6$ \sp
&
$A_{3/2} = {\bf b_0}   + b_1  \tilde q^2 + b_2  \tilde q^4 + b_3  \tilde q^6$ \sp
& 
$S_{1/2} = {\bf c_0} \tilde q + c_1  \tilde q^3 + c_2  \tilde q^5 + c_3 \tilde q^7$ \\[.1cm]
\hline
\hline
\end{tabular}
\end{center}
\caption{\footnotesize
Parametrizations of the helicity amplitudes for the cases $\frac{1}{2}^\pm$ 
and $\frac{3}{2}^\pm$.}
\label{tabAmps}
\end{table*}

\section{Analytic extension of the helicity amplitudes
to the timelike region}
\label{secAnalytic}

In the region $ - (M_R -M_N)^2 <Q^2  < Q_P^2$,  we 
consider a formal representation of the helicity amplitudes 
in terms of the variable $|{\bf q}|$ defined by Eq.~(\ref{eqq2}),
in order to parametrize the leading-order dependence 
of the amplitudes near the pseudothreshold.
The connection between $Q^2$ and $|{\bf q}|$ can be obtained 
by the inversion of the relation (\ref{eqq2}):
\ba
Q^2 = -(M_R -M_N)^2 + 2 M_R \left[ \sqrt{M_N^2 + |{\bf q}|^2} - M_N\right].  
\nonumber \\
\label{eqQ2a}
\ea
 
More specifically, we represent the helicity amplitudes 
using an expansion  in powers of $ |{\bf q}|^2$ in the form 
\ba
A= |{\bf q}|^n ( \alpha_0 + \alpha_1  |{\bf q}|^2 + 
\alpha_2  |{\bf q}|^4 + \alpha_3  |{\bf q}|^6),
\label{eqPowerQT}
\ea
for the cases $n=0,1,2$.
The representation (\ref{eqPowerQT})  
is consistent with all the amplitudes from Table~\ref{tableST1}.
The coefficients $\alpha_0$,  $\alpha_1$,  
 $\alpha_2$ and    $\alpha_3$ are determined 
by the connection with the $Q^2 > Q_P^2$ region 
and the continuity conditions associated 
with the amplitudes, the first and the second derivatives of the amplitudes,
as discussed next.
In some cases we use also the continuity of the third derivative.

Note that in Eq.~(\ref{eqPowerQT}) there 
are no odd powers of $|{\bf q}|$ in the second factor.
This representation is motivated by 
the relation between the derivative 
in $|{\bf q}|$ and $Q^2$, from where we can conclude 
that the odd terms vanish near 
the pseudothreshold\footnote{The relation between a derivative 
of a function  $F$
in  $|{\bf q}|$  and $Q^2$ is
\ba
\frac{d F}{d |{\bf q}|} =
\frac{4 M_R^2   |{\bf q}|}{ M_R^2 + M_N^2 + Q^2} \frac{d F}{d Q^2}. 
\nonumber
\ea
If $F$ is a regular function, (no singularities at  $|{\bf q}|=0$),
one concludes that $\frac{d F}{d |{\bf q}|}$ vanishes at the pseudothreshold.}.

Instead of the variable $|{\bf q}|^2$, one can use 
the dimensionless variable
\ba
\tilde q^2 = \frac{|{\bf q}|^2}{M_R^2}.
\label{eqQT}
\ea  
For simplicity, we define also $\tilde q = \sqrt{\tilde q^2}$.
Using the new notation, we can parametrize
the amplitudes for the $\frac{1}{2}^\pm$ and  $\frac{3}{2}^\pm$ 
resonances in the form 
\ba
A=
\tilde q^n ( \alpha_0 + \alpha_1  \tilde q^2 + 
\alpha_2  \tilde q^4 + \alpha_3  \tilde q^6),
\label{eqPowerQT2}
\ea
where all the coefficients $\alpha_l$  ($l=0,1,2,3$) 
have the same dimensions, the dimension of the 
helicity amplitudes (GeV$^{-1/2}$).
The explicit parametrizations for the resonances
under study are in Table~\ref{tabAmps}.

In Table~\ref{tabAmps}, we use $a_l$, $b_l$ and $c_l$ 
($l=0,1,2,3$) to represent the coefficients 
of the amplitudes $A_{1/2}$, $A_{3/2}$ and $S_{1/2}$, respectively.
The effect of the factor $\tilde q^n$ in Eq.~(\ref{eqPowerQT2})
can be taken into 
account by redefining $A$ as $\frac{A}{\tilde q}$ or $\frac{A}{\tilde q^2}$.

The coefficients  $\alpha_l$  ($l=0,1,2,3$) 
are determined based on the correlations between 
the amplitudes (pseudothreshold conditions) 
and the continuity conditions for $A$ and  
their derivatives, characterized by the moments $A^{(k)}$ ($k=0,1,2,3$).
There are two cases to be considered:
\begin{enumerate}
\item
The amplitude is independent of the pseudothreshold conditions.
\item
The amplitude is correlated with another amplitude.
\end{enumerate}

In the first case, we can determine all coefficients 
using the continuity of $A$ and the first three derivatives
at the point $Q^2= Q_P^2$ to fix all the coefficients.
The last coefficient $\alpha_3$ is in this case 
determined by $A^{(3)}$ (third derivative of $A$).

In the second case we use the correlation condition 
to fix the first coefficient ($\alpha_0$) 
and the remaining coefficients are determined 
by the continuity of the amplitude and the 
first two derivatives at the point $Q^2= Q_P^2$.

The explicit expressions for the two cases 
are presented in Appendix~\ref{appendix-alpha}
for the function (\ref{eqPowerQT2}) for the case $n=0$.
The other cases can be obtained using the corresponding results for 
$\frac{A}{\tilde q}$ or $\frac{A}{\tilde q^2}$. 
The relations between the coefficients $\alpha_l$
and the moments of the expansion in $Q^2$ from Eq.~(\ref{eqAmp1})
are presented in Appendix~\ref{appDerivatives}.

Since the longitudinal amplitude ($S_{1/2}$) is, 
in general, poorly constrained near $Q^2=0$, because 
there are no measurements at  $Q^2=0$, 
in our calculations we choose to fix the 
coefficients of $S_{1/2}$ using the correlations 
with the transverse amplitudes.

The coefficients of the transverse amplitudes 
($A_{1/2}$ and $A_{3/2}$) can, in principle, be considered independent of 
the pseudothreshold conditions.
In those conditions all the coefficients 
can be determined using the continuity of the amplitude 
and the first three derivatives, as mentioned above.
The states $\frac{3}{2}^-$ are the exception to this 
role because the two transverse amplitudes are also correlated,
as shown in the second column in Table~\ref{tableST1}.

We now consider the different $J^P$ states.

\subsection{$\frac{1}{2}^+$ states}

For the $\frac{1}{2}^+$ states there are no special 
constraints at the pseudothreshold, 
except for the forms
\ba
A_{1/2} \propto |{\bf q}|, 
\hspace{.6cm} 
S_{1/2} \propto |{\bf q}|^2,
\label{eqP11-1}
\ea
presented in Table~\ref{tableST1}.

In the present case, 
we can determine all coefficients 
of the amplitudes $A_{1/2}$ and $S_{1/2}$
demanding the continuity of the amplitudes and 
the first three derivatives independently,
since those amplitudes are uncorrelated (case 1).

From the parametrizations from Ref.~\cite{Jlab-fits},
there are two resonances to be taken into account:
$N(1440)$ (Roper) and $N(1710)$.

\subsection{$\frac{1}{2}^-$ states}

The $\frac{1}{2}^-$ states are characterized 
by the relation~\cite{Devenish76,Siegert1}
\ba
A_{1/2} = \sqrt{2} (M_R -M_N)  \frac{S_{1/2}}{|{\bf q}|}.
\label{eqS11-1}
\ea
The previous relation is equivalent 
to Eq.~(\ref{eqST}), since $E \propto A_{1/2}$.

An alternative view of the pseudothreshold 
constraints is obtained when 
we consider the Dirac ($F_1$) and Pauli ($F_2$) form factors.
In that case we can write, near the pseudothreshold
$A_{1/2} = 2 b \tilde F_1$ 
and \mbox{$S_{1/2} = \sqrt{2} b \frac{|{\bf q}|}{M_R -M_N} \tilde F_1$,}
where $\tilde F_1 = F_1 + \frac{M_R -M_N}{M_R + M_N} F_2$
and $b \propto \sqrt{Q_+^2}$ is a known function~\cite{Siegert1}\footnote{The 
Dirac form factor is related 
with $G_1$ from Eqs.~(\ref{eqRoper}) by $F_1 = \tau G_1$.
In addition $G_2= F_2$}. 
Check Ref.~\cite{Siegert1} for more details.

According to the parametrizations from 
Table~\ref{tabAmps}, the condition  (\ref{eqS11-1})  
corresponds to  
\ba
\sqrt{2}
\frac{M_R -M_N}{M_R}c_0 = a_0.
\label{eqS11-2}
\ea
The bold variable $\mathbf{c_0}$ 
in Table~\ref{tabAmps} indicates that $c_0$ 
is fixed by (\ref{eqS11-2}).

The amplitude $A_{1/2}$ can be considered independent 
with the coefficients determined by the continuity conditions  
of the amplitude and the first three derivatives (case 1).
The coefficient $c_0$ is determined by the value of $a_0$.
The remaining coefficients $c_l$ ($l=1,2,3$) 
are determined by the continuity conditions associated with the 
amplitude $S_{1/2}$ and the first 
two derivatives (case 2).

This formalism can be used for the states 
$N(1535)$, $N(1650)$ and $\Delta(1620)$.

\subsection{$\frac{3}{2}^+$ states}

The $\sfrac{3}{2}^+$ are constrained by the condition 
between electric and Coulomb Jones-Scadron 
form factors~\cite{Devenish76,Jones73}
\ba
G_E = \frac{M_R -M_N}{2 M_R} G_C.
\label{eqD33}
\ea
When expressed in terms of helicity amplitudes, 
we obtain the relations 
$G_E = F_+ \left( A_{1/2} - \sfrac{1}{\sqrt{3}} A_{3/2}\right)$
and $\frac{|{\bf q}|}{2 M_R} G_C = F_+ \sqrt{2} S_{1/2}$,
where $F_+ \propto 1/|{\bf q}|$ is a kinematic factor.
One obtains then the relation 
between amplitudes~\cite{Jones73}
\ba
A_{1/2} - \frac{1}{\sqrt{3}} A_{3/2}
= \sqrt{2} \frac{M_R-M_N}{|{\bf q}|} S_{1/2}.
\ea
According to the parametrizations 
from Table~\ref{tabAmps},
the previous equation is equivalent to  
\ba
\sqrt{2}
\frac{M_R -M_N}{M_R} c_0 = a_0 - \frac{b_0}{\sqrt{3}}.
\label{eqP33-1}
\ea

An additional consequence of the relations 
$A_{1/2} \propto |{\bf q}|$ and  $A_{3/2} \propto |{\bf q}|$ is that 
\ba
G_M \propto 
 (A_{1/2} - \sfrac{1}{\sqrt{3}} A_{3/2})/|{\bf q}|
\propto 1, 
\ea
near the pseudothreshold, meaning 
that $G_M$ is finite when $|{\bf q}| \to 0$~\cite{Siegert2,Compton}.

In the present case  the amplitudes $A_{1/2}$ and $A_{3/2}$ 
are uncorrelated and all coefficients can be determined independently 
using the continuity of the amplitudes and the first three derivatives (case 1).
The coefficient $c_0$ is determined afterwards  
using Eq.~(\ref{eqP33-1}).
The remaining coefficients are determined by 
the continuity of the function and the 
first two derivatives at the point $Q^2= Q_P^2$ (case 2).

This formalism can be applied to the resonances $\Delta(1232)$ and $N(1720)$.

\subsection{$\frac{3}{2}^-$ states}

Contrary to the previous cases there are 
two conditions at the 
pseudothreshold:~\cite{Siegert2,Compton,Devenish76,Aznauryan12b}:
\ba
& &
G_E = -\frac{M_R - M_N}{M_N} G_C, \\
& &
G_M \propto |{\bf q}|^2. 
\ea
One has then two constraints for the three form factors.
When expressed in terms of the helicity amplitudes, we obtain:
\ba
& &
A_{1/2} +  \sqrt{3} A_{3/2}
= - 2 \sqrt{2} \frac{M_R-M_N}{|{\bf q}|} S_{1/2}, \\
& &
A_{1/2} - \frac{1}{\sqrt{3}} A_{3/2} \propto |{\bf q}|^2.
\ea
Based on the parametrization from Table~\ref{tabAmps},
we derive the relations between coefficients
\ba
& &
b_0= \sqrt{3} a_0, 
\label{eqD13-1}
\\
& &
-\frac{M_R -M_N}{\sqrt{2}M_R}c_0 = a_0.
\label{eqD13-2}
\ea

In the case of the resonances $\frac{3}{2}^-$ 
one has then two constraints for the three amplitudes 
expressed by Eqs.~(\ref{eqD13-1}) and (\ref{eqD13-2}).
Since $a_0$ appears in both conditions,
we use the continuity of $A_{1/2}$ and 
the first three derivatives to fix the coefficients $a_l$ first (case 1).
After that, we fix the coefficients $b_0$ and $c_0$
using  Eqs.~(\ref{eqD13-1}) and (\ref{eqD13-2})
and determine the remaining coefficients 
through the continuity of $A_{3/2}$ and $S_{1/2}$ 
and the first two derivatives (case 2).

The present formalism can be applied 
to the resonances $N(1520)$ and  $\Delta(1700)$.

\begin{figure*}[t]
%\vspace{.5cm}
\centerline{\vspace{0.5cm}  }
\centerline{
\mbox{
\includegraphics[width=2.5in]{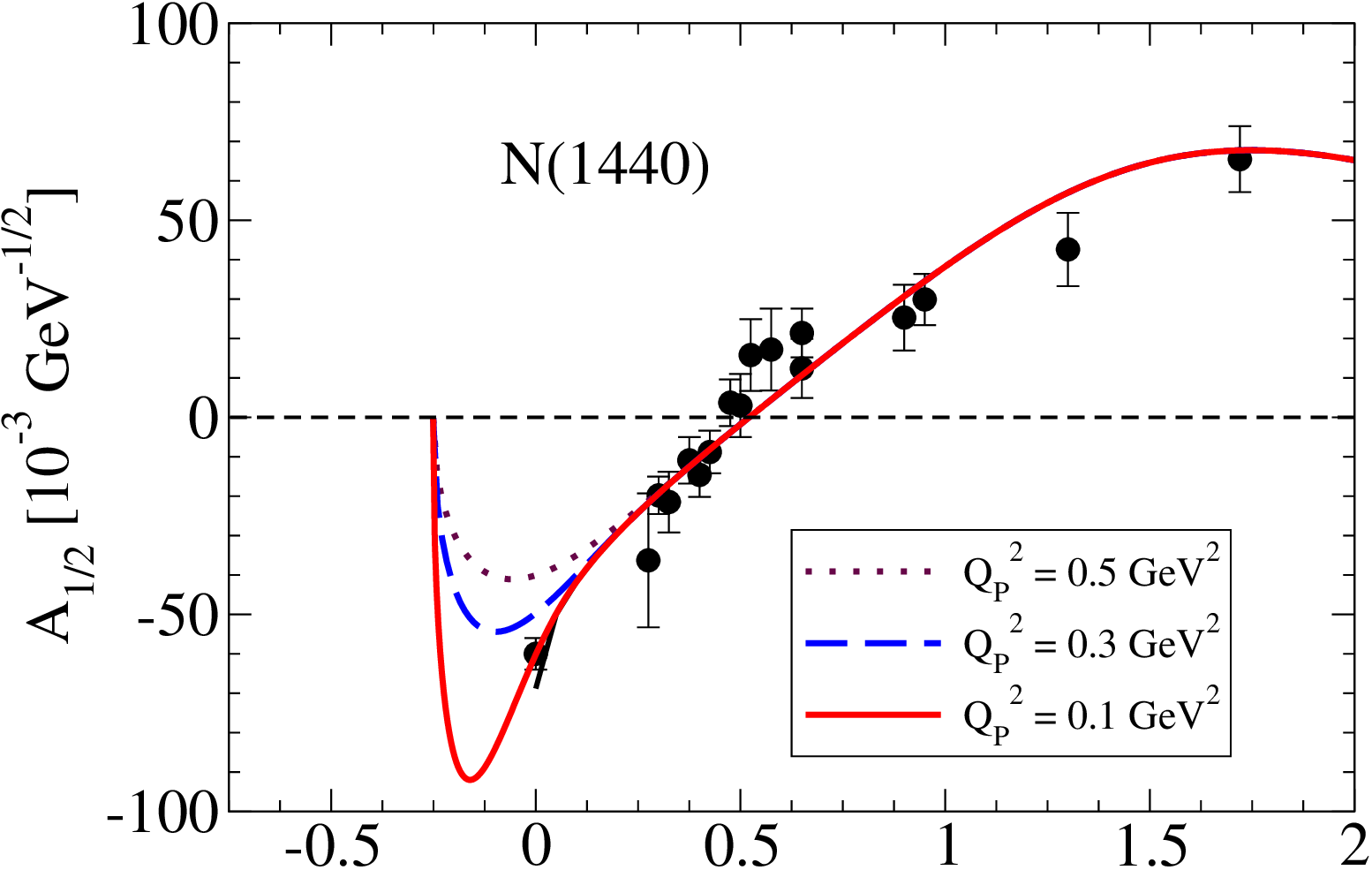} \hspace{.6cm}
\includegraphics[width=2.5in]{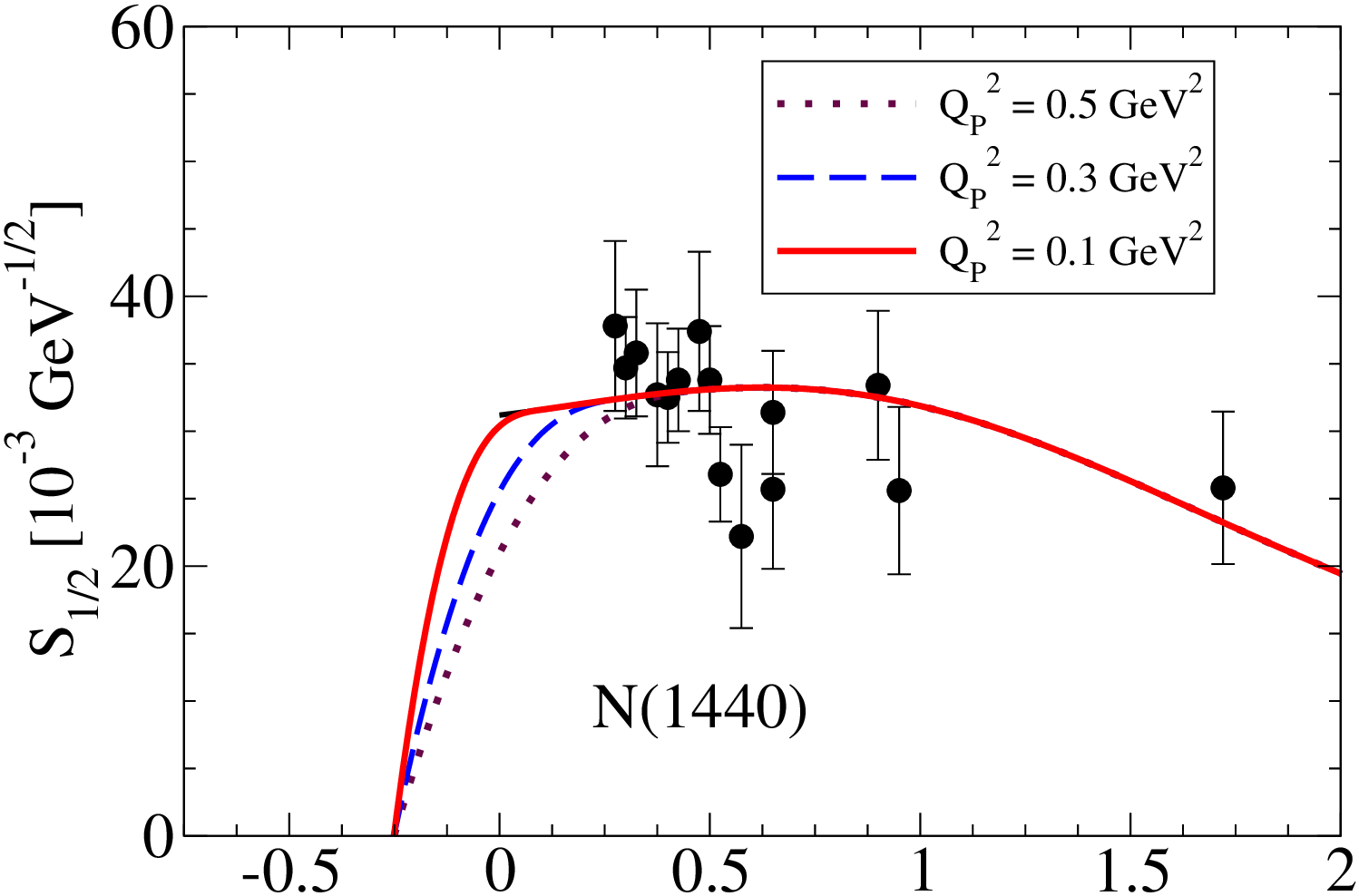} }}
\vspace{.7cm}
\centerline{
\mbox{
\includegraphics[width=2.5in]{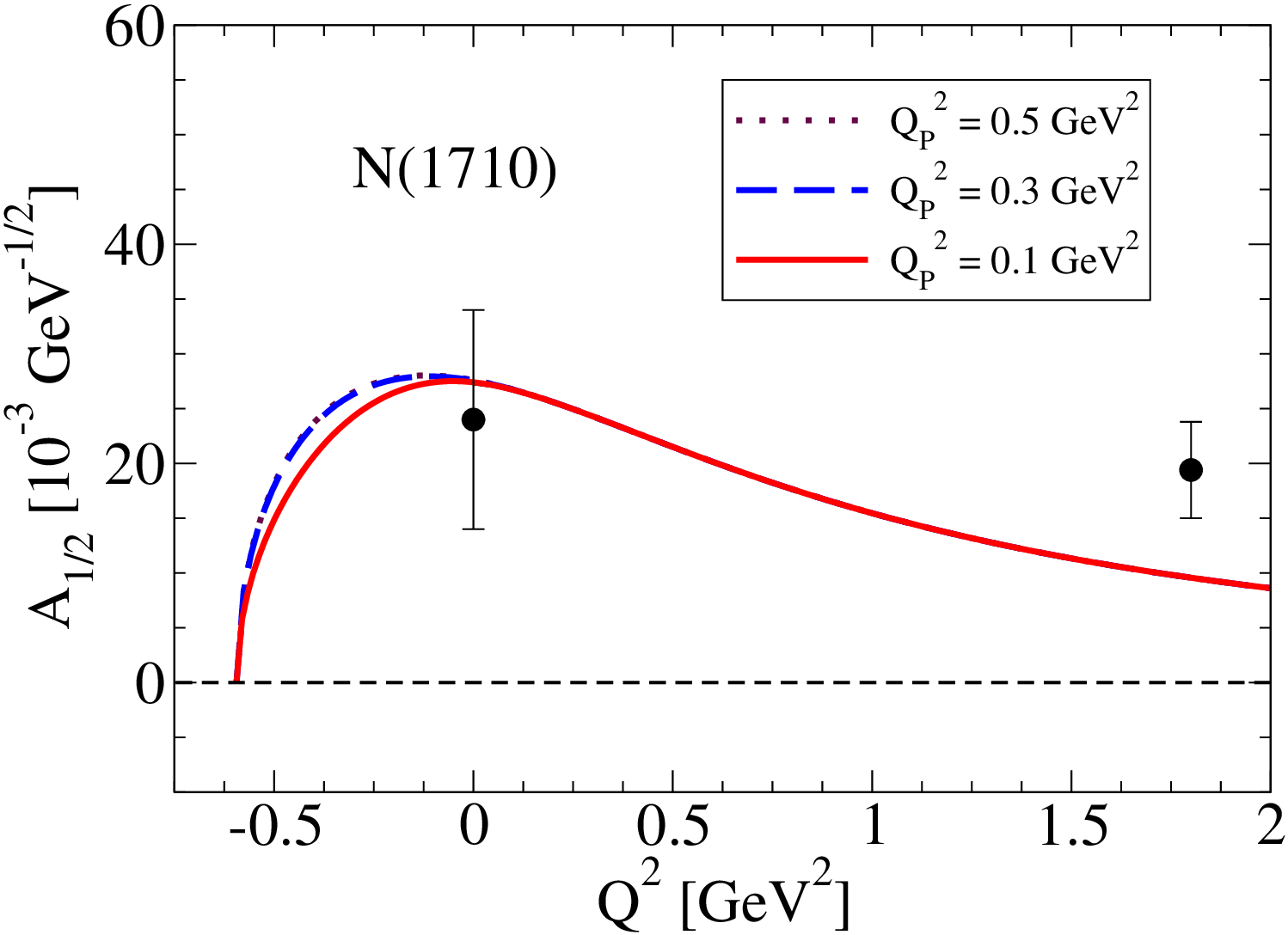} \hspace{.6cm}
\includegraphics[width=2.5in]{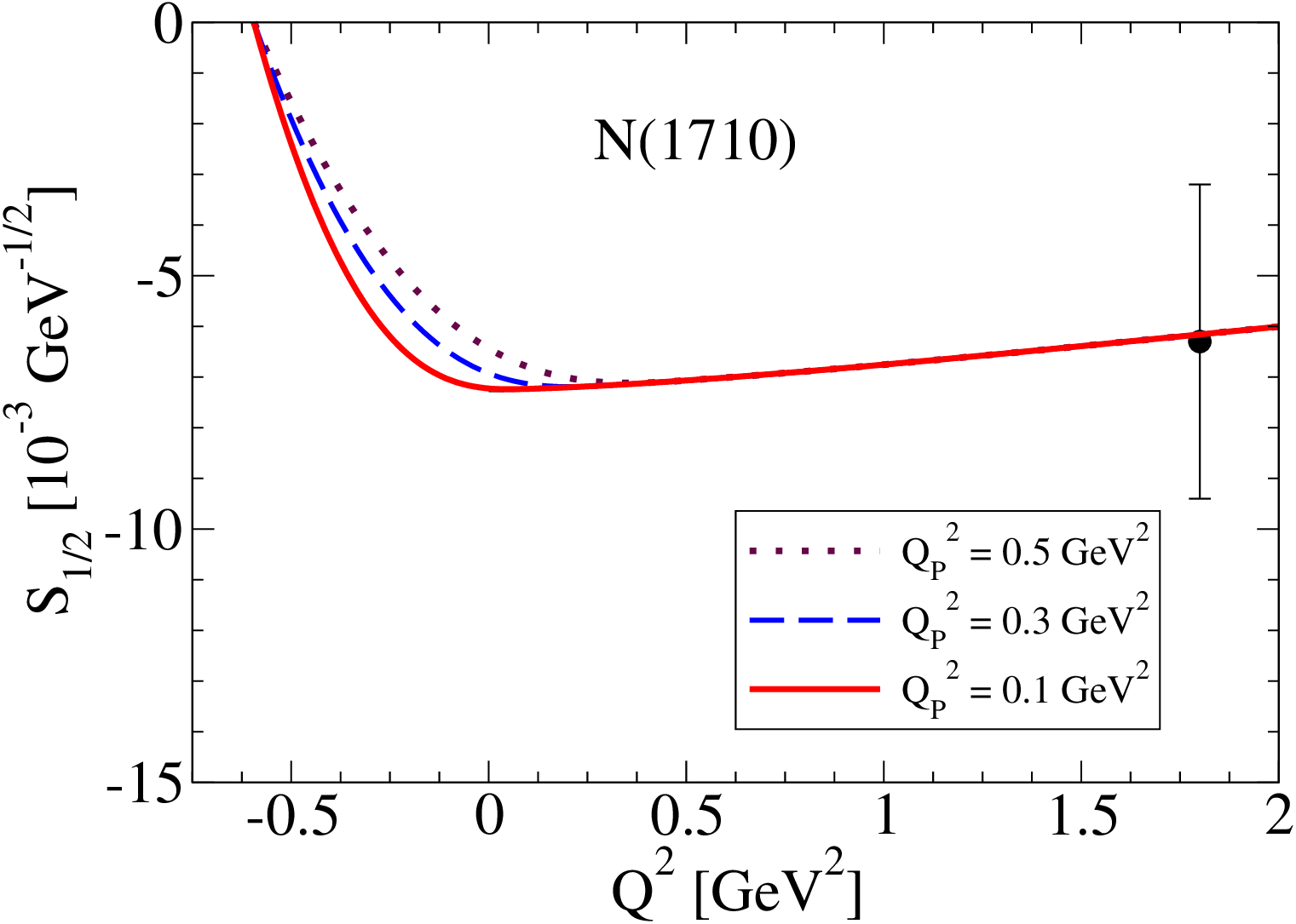} }}
\caption{\footnotesize{$\gamma^\ast N \to N^\ast 
\left(\frac{1}{2}^+ \right)$ 
transition amplitudes. $N^\ast= N(1440)$, $N(1710)$.
Data for $N(1440)$ from Refs.~\cite{PDG14,CLAS09,CLAS12,Mokeev16}.
The lowest point for $S_{1/2}$ are from MAMI~\cite{Stajner17}.
Data for $N(1710)$ from Refs~\cite{PDG14,Park15}
(three data points for $Q^2 > 2$ GeV$^2$ are not shown).}}
%\vspace{-1cm}
\label{fig-onehalfp}
\end{figure*}

\begin{figure*}[t]
%\vspace{.5cm}
\centerline{\vspace{0.5cm}  }
\centerline{
\mbox{
\includegraphics[width=2.5in]{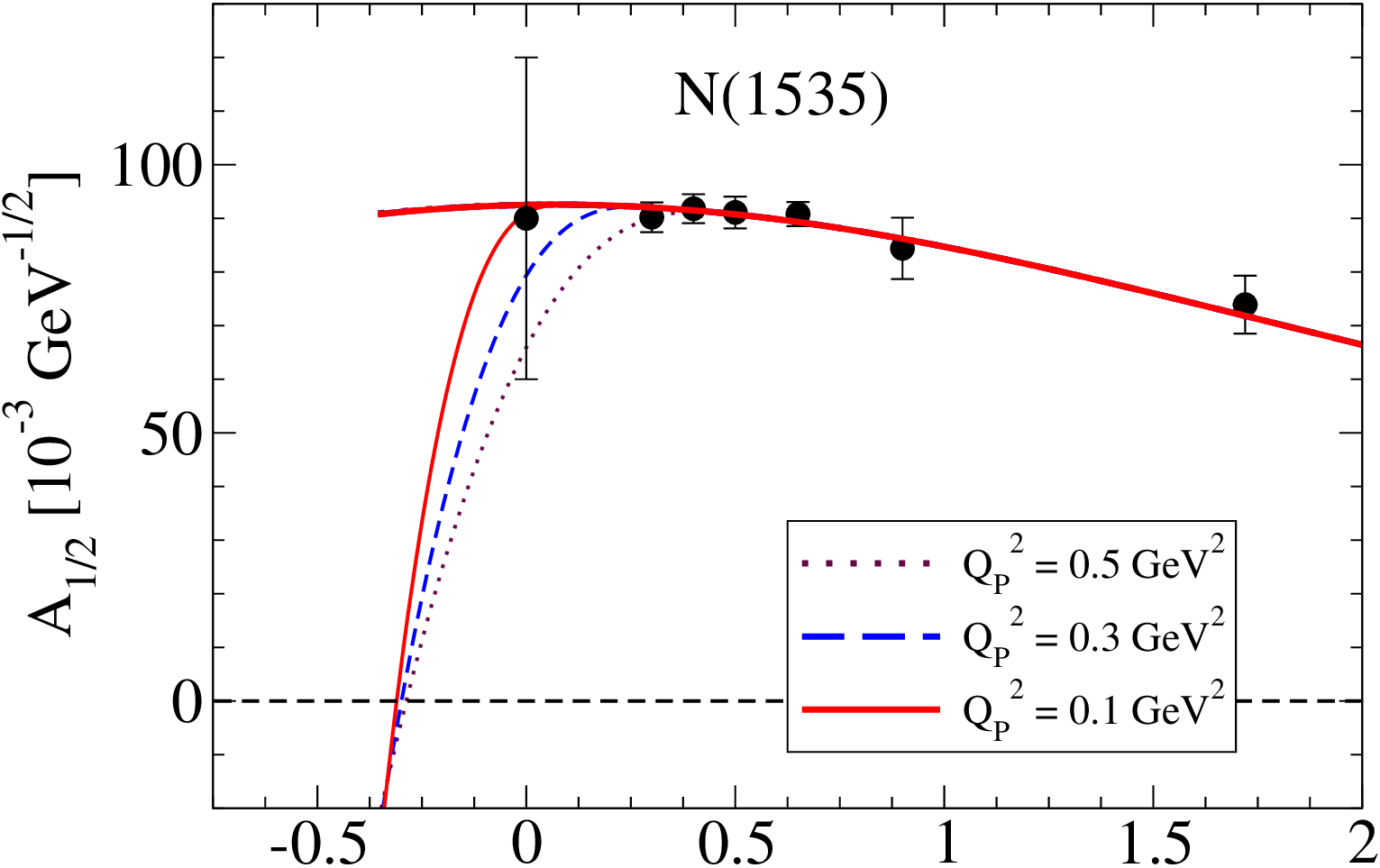} \hspace{.6cm}
\includegraphics[width=2.5in]{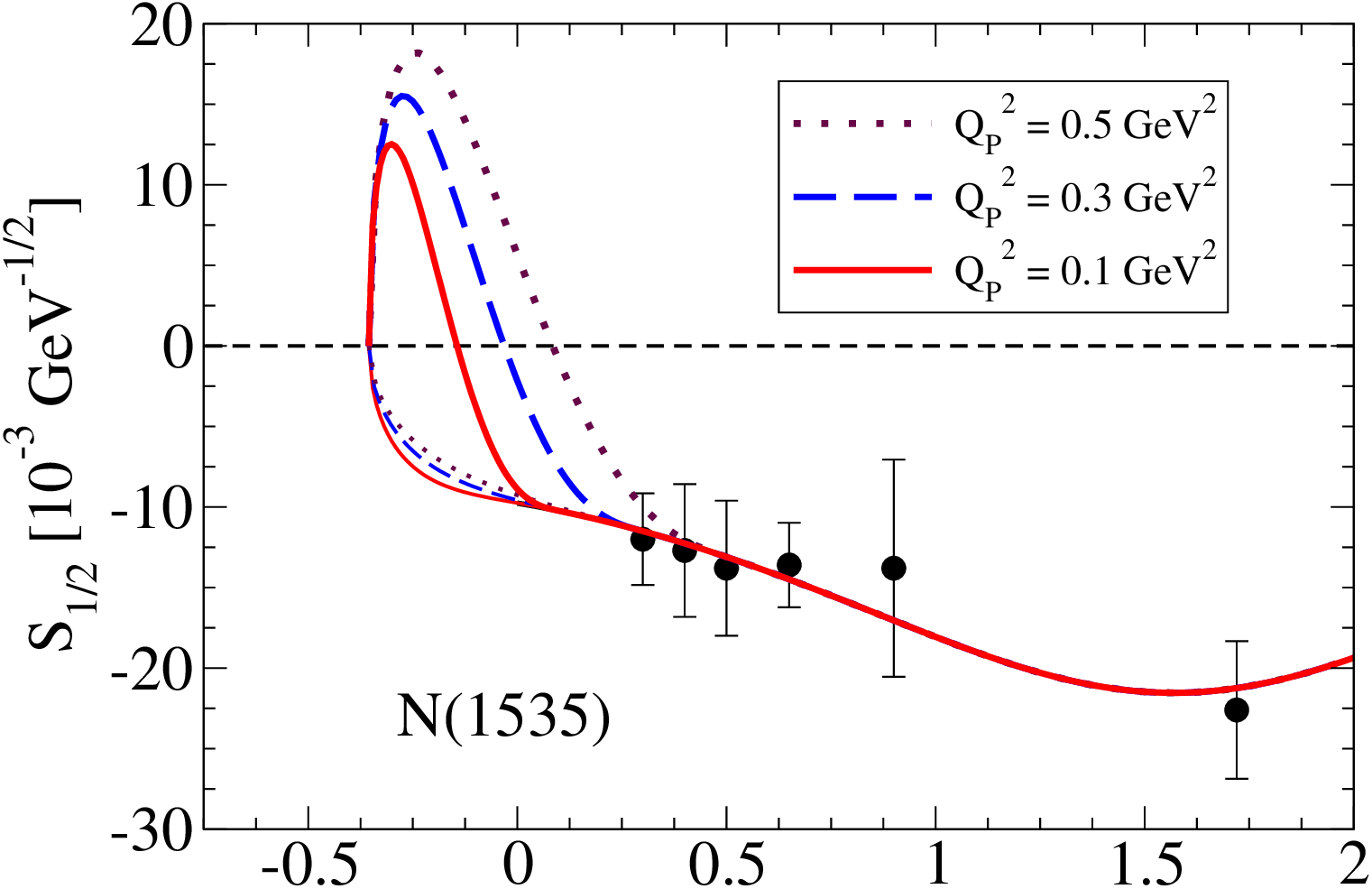} }}
\vspace{.7cm}
\centerline{
\mbox{
\includegraphics[width=2.5in]{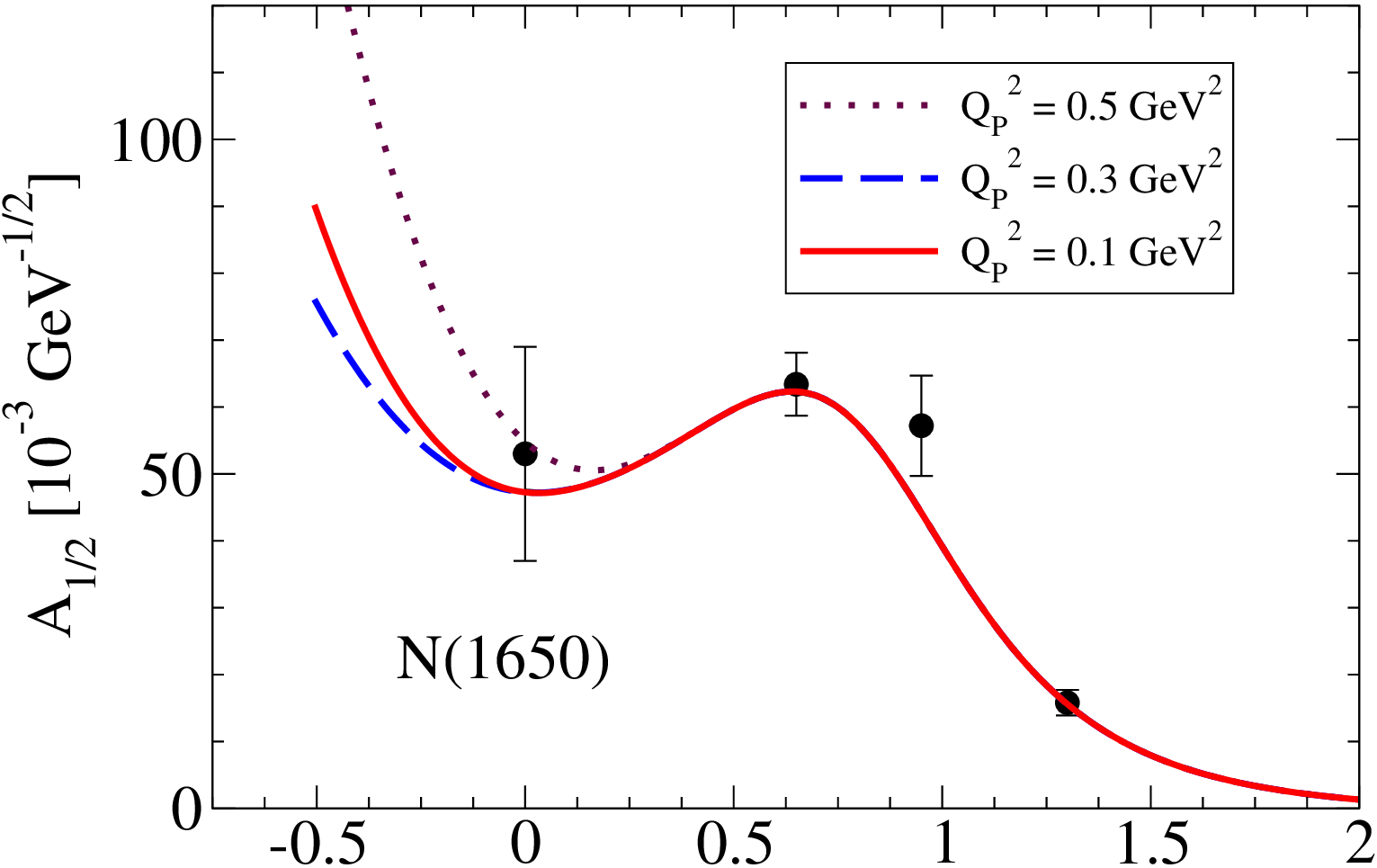} \hspace{.6cm}
\includegraphics[width=2.5in]{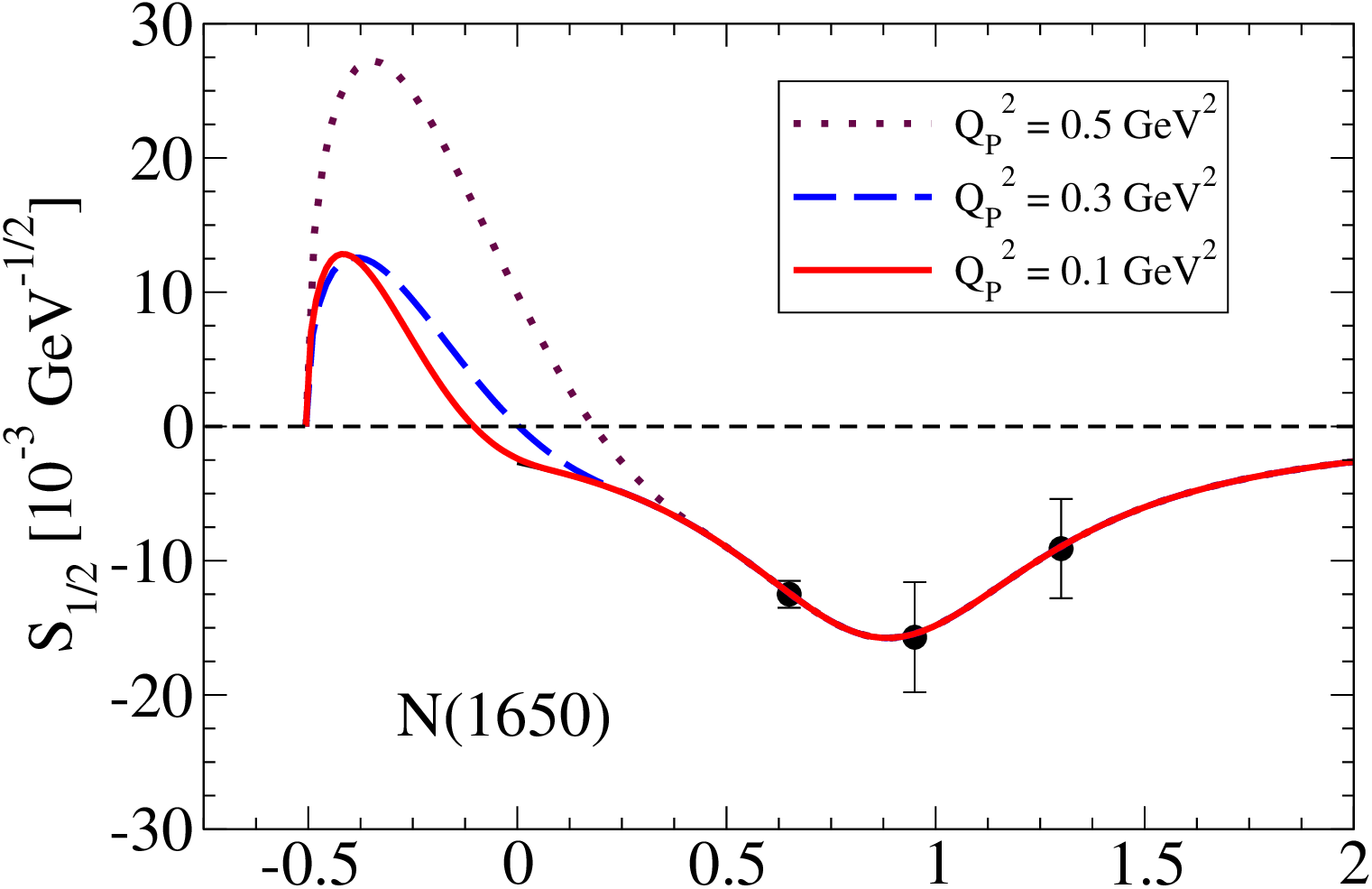} }}
\vspace{.73cm}
\centerline{
\mbox{
\includegraphics[width=2.5in]{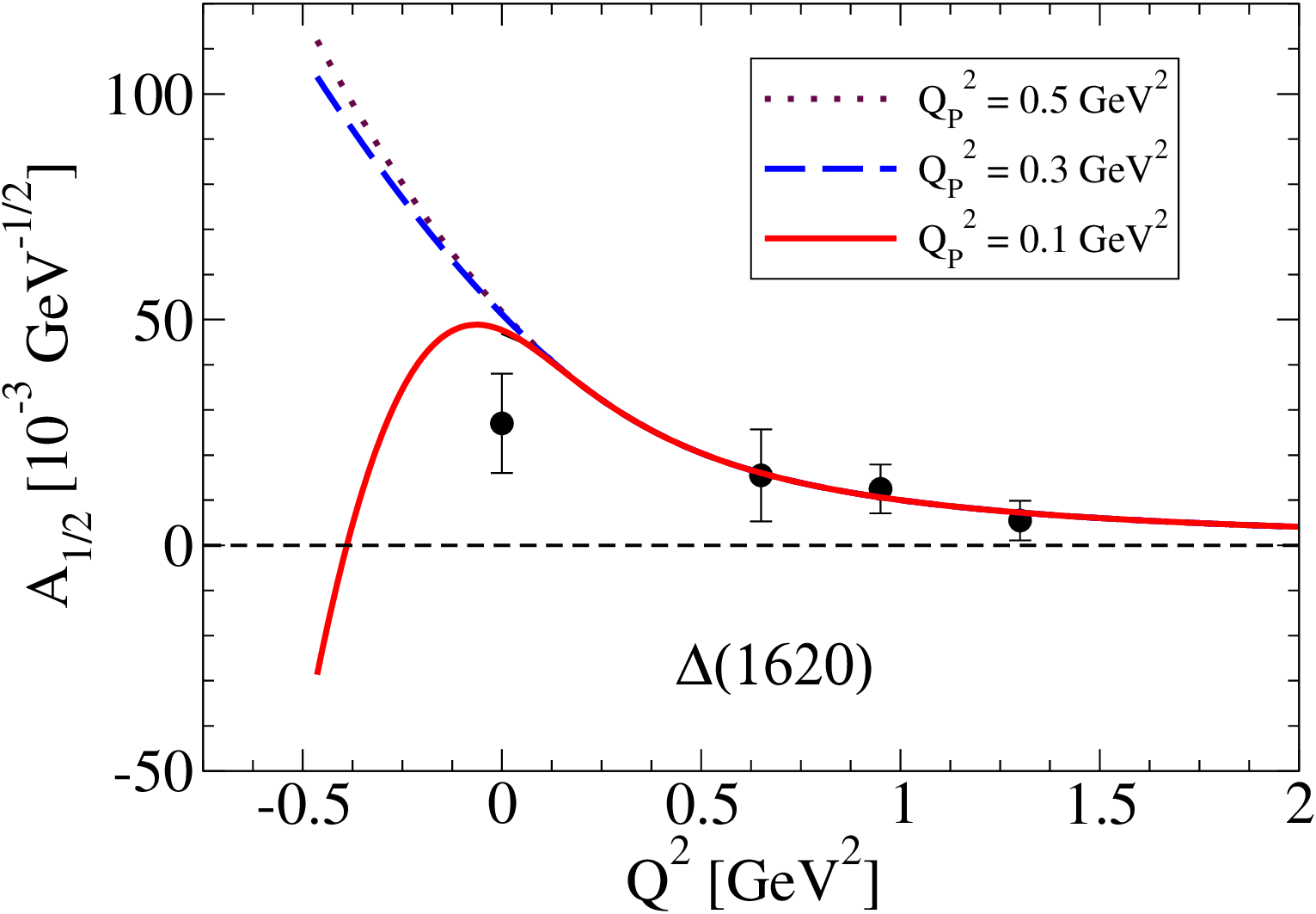} \hspace{.6cm}
\includegraphics[width=2.5in]{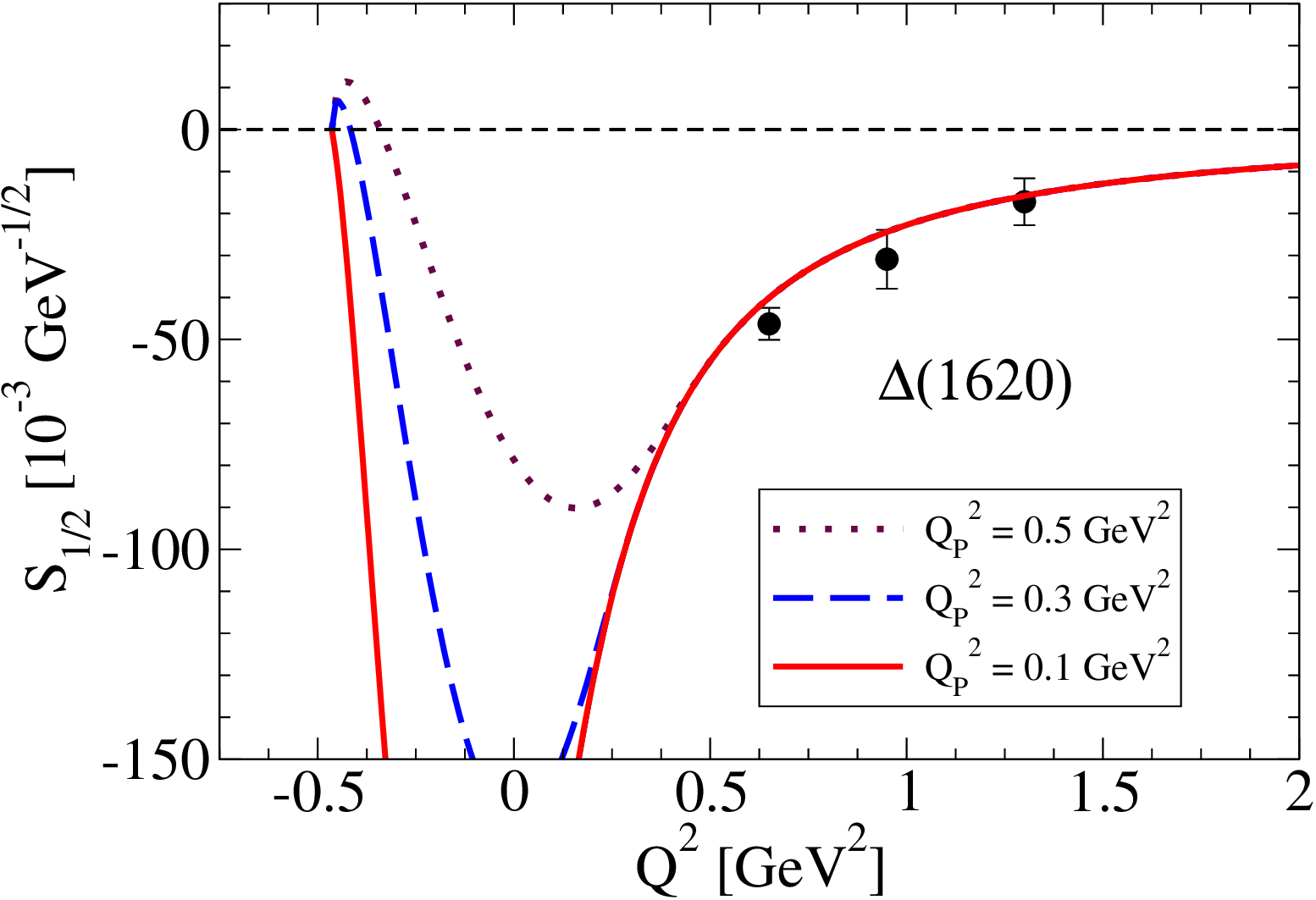} }}
\caption{\footnotesize{$\gamma^\ast N \to N^\ast 
\left(\frac{1}{2}^- \right)$ 
transition amplitudes. $N^\ast= N(1535)$, $N(1650)$,
$\Delta(1620)$. 
The tick lines for the $N(1535)$ 
describe alternative parametrizations discussed in 
the main text.
Data for $N(1535)$ from Ref.~\cite{PDG14,CLAS09}.
Data for $N(1650)$ from Ref.~\cite{PDG14,Mokeev14}.
Data for $\Delta(1620)$ from Ref.~\cite{PDG14,Mokeev16}.}}
%\vspace{-1cm}
\label{fig-onehalfm}
\end{figure*}

%\clearpage

\section{Numerical results}
\label{secResults}

We present now the numerical results 
associated with different $\frac{1}{2}^\pm$ and  $\frac{3}{2}^\pm$ states.

The formalism discussed in the previous section is 
general, and can be applied to any regular 
set of parametrizations of the amplitudes,
in the spacelike region 
(the amplitudes and the first derivatives are continuous).

To exemplify the method, we consider 
the  empirical parametrizations from the Jefferson Lab 
from  Ref.~\cite{Jlab-fits}.
Those parametrizations are based on simple rational expressions,
in general, are  valid in the range $Q^2=0.5$--5 GeV$^2$, 
and provide a fair description of the available data, 
in particular the CLAS data.

To study the sensibility of the analytic extensions 
to the transition point between the original parametrization 
and our analytic extension  ($Q_P^2$), we consider the values 
$Q_P^2=0.1$, 0.3 and 0.5 GeV$^2$.

Our analytic extensions are also compared 
directly with the data, particularly
the database from Ref.~\cite{MokeevDatabase}.
We recall that the $N^\ast$ helicity amplitude
data are, in general, scarce, 
except for the states $\Delta(1232)\frac{3}{2}^+$, $N(1440)\frac{1}{2}^+$, 
$N(1520)\frac{3}{2}^-$ and  $N(1535)\frac{1}{2}^-$.
The data for most of the other states are  
incomplete, since they are restricted mainly to 
the transverse amplitudes and also restricted to 
three or five data points.
For these reasons, we select, in particular, data from CLAS:
The CLAS data are complete and cover a wide range in $Q^2$.

Another limitation of the data is that there is 
a lack of data in the region $Q^2=0$--0.3 GeV$^2$.
As a consequence, the analytic extrapolations 
for the low-$Q^2$ region are very sensitive to the available data, 
leading to ambiguities in the extrapolations to the $Q^2 < 0$ region. 
These effects are sometimes amplified by the correlation between the 
amplitudes near the pseudothreshold.   

Of particular importance is the experimental 
determination of the transverse amplitudes at the photon point ($Q^2=0$).
Different groups provide very different estimates for those amplitudes.
In some some cases, the  Particle Data Group (PDG) summarizes the results 
using an interval with a large window of variation.

\begin{figure*}[t]
%\centerline{\vspace{0.5cm}  }
%\vspace{0.5cm}
\centerline{\vspace{0.5cm}  }
\centerline{
\mbox{
\includegraphics[width=2.1in]{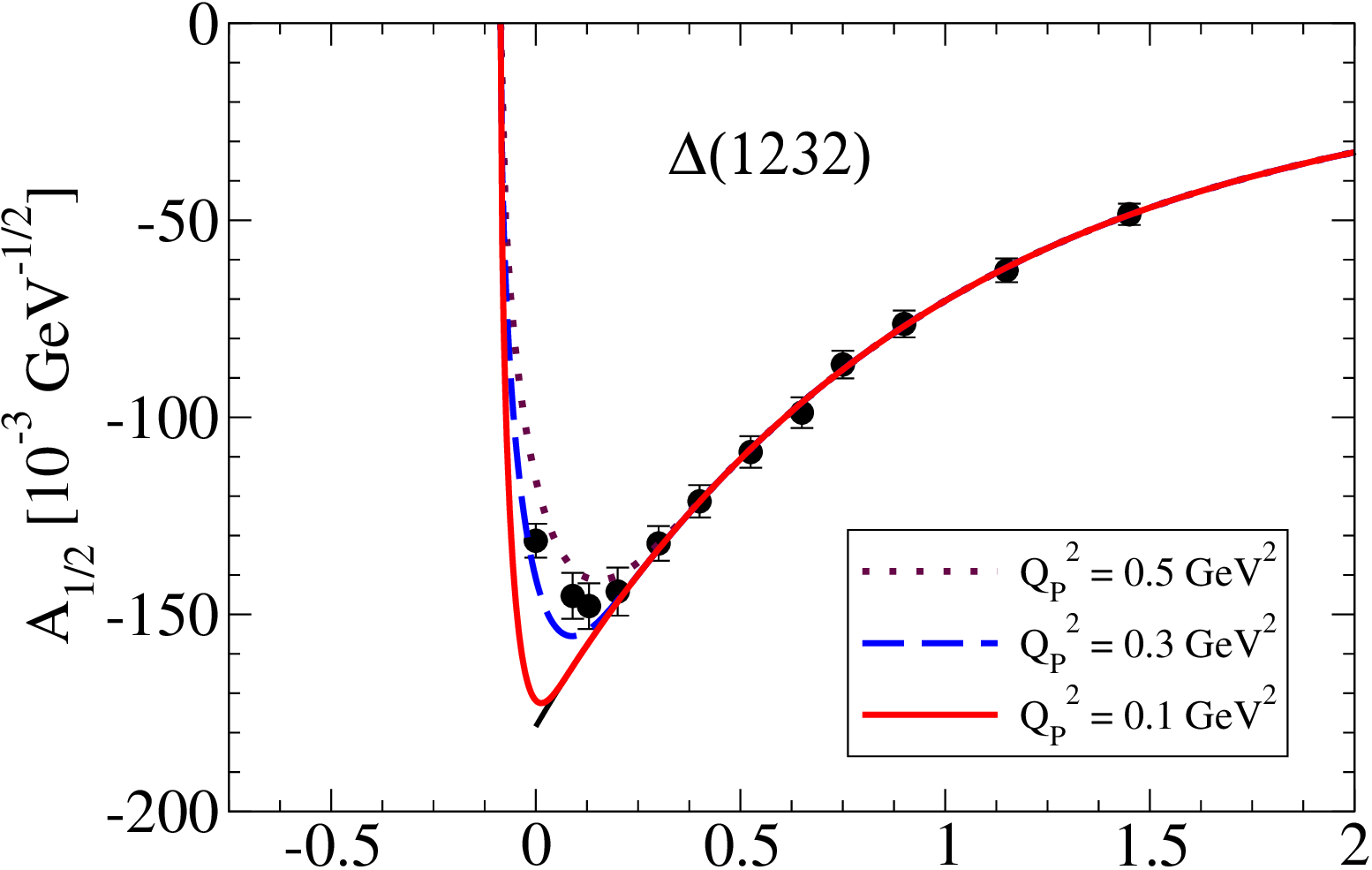} \hspace{.3cm}
\includegraphics[width=2.1in]{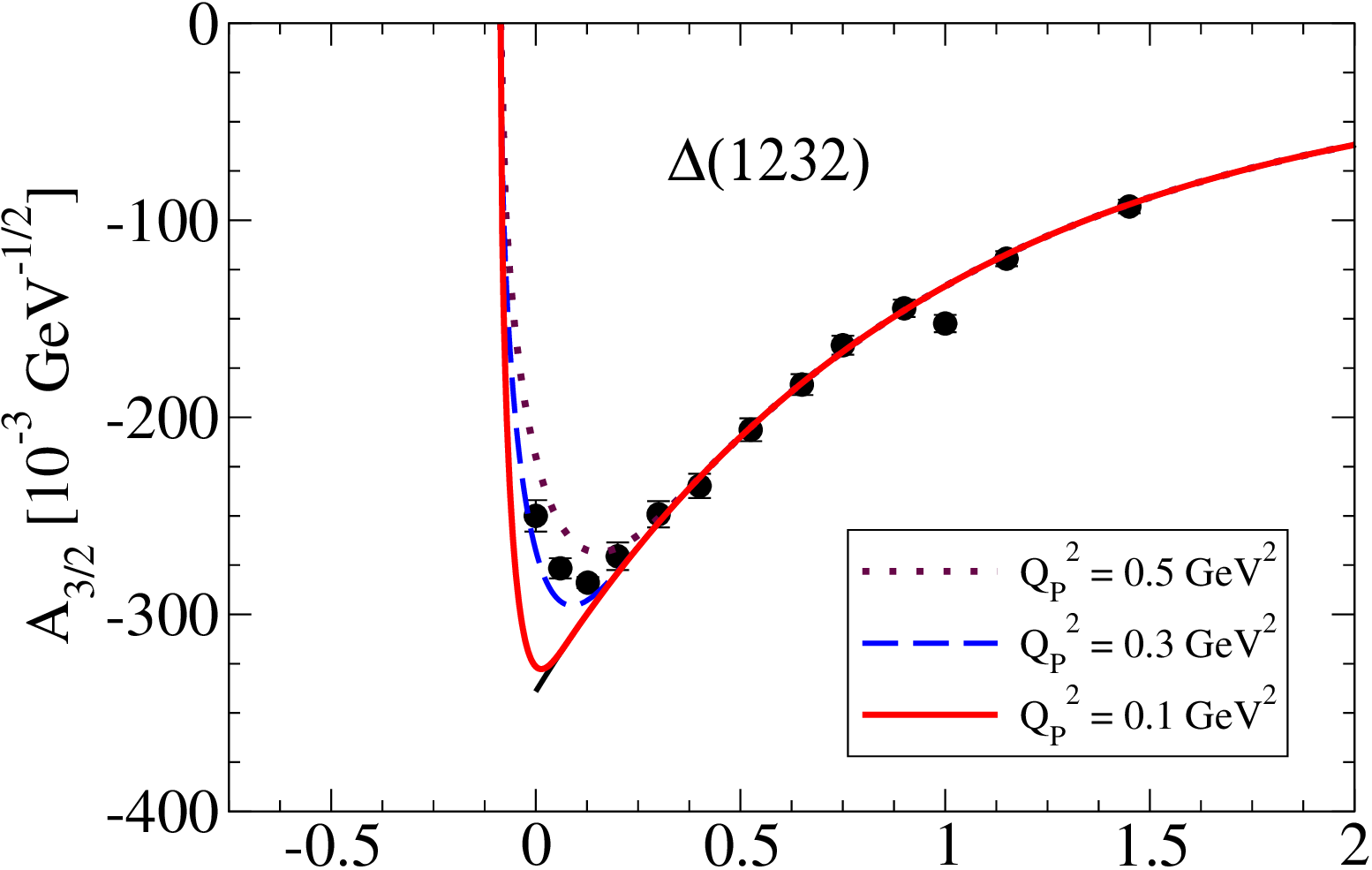}  \hspace{.3cm}
\includegraphics[width=2.1in]{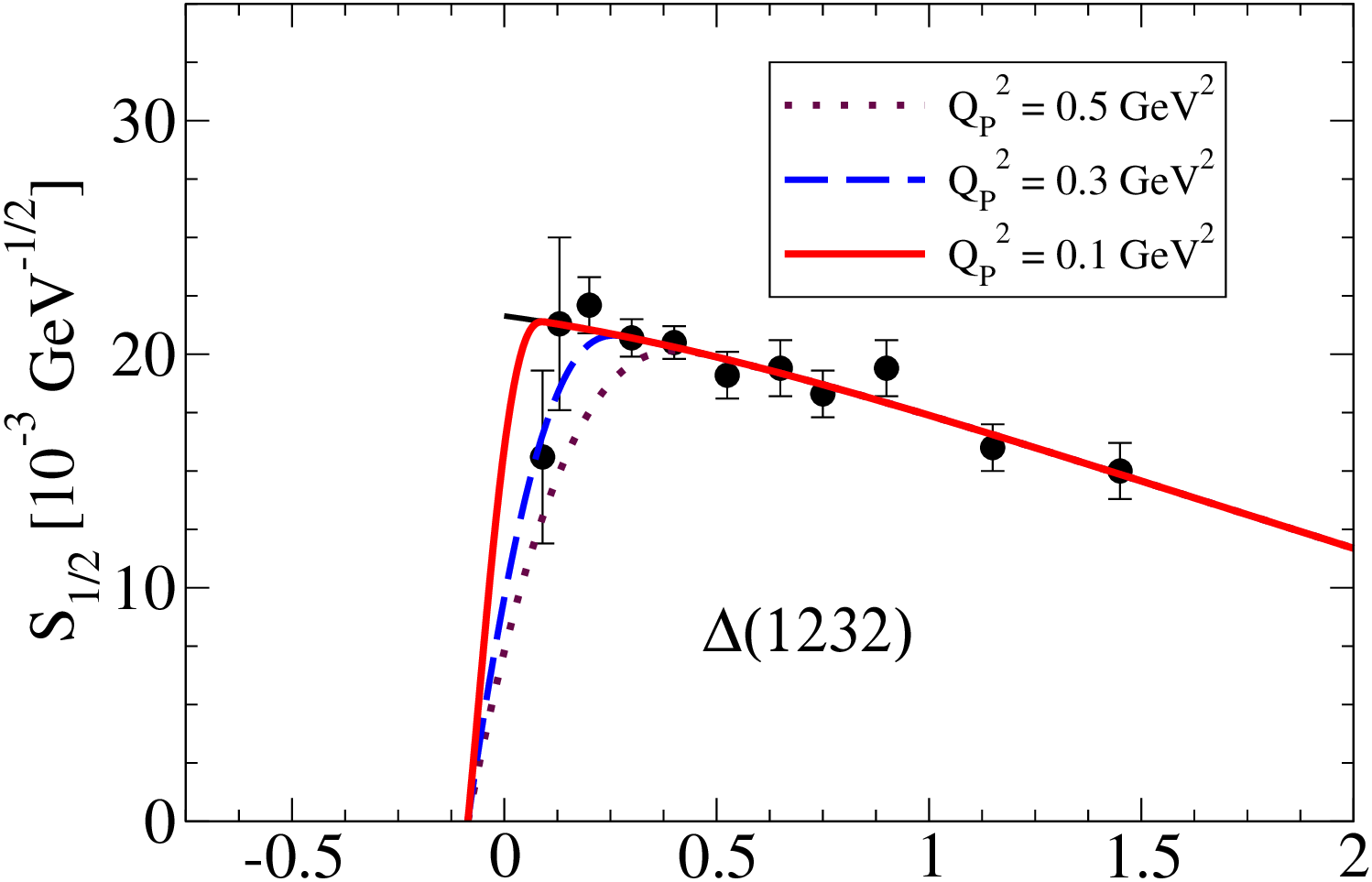} }}
\vspace{.7cm}
\centerline{
\mbox{
\includegraphics[width=2.1in]{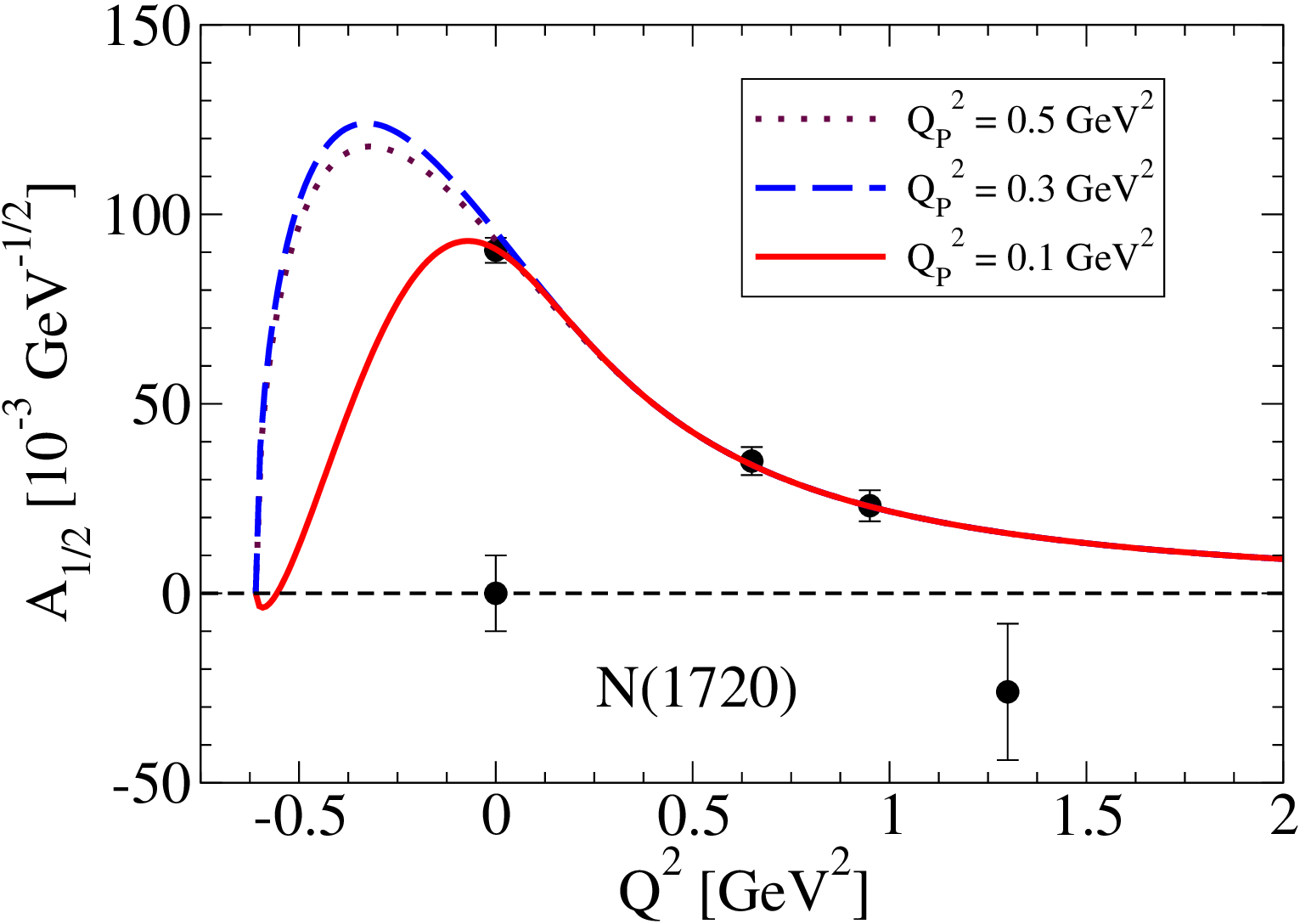} \hspace{.3cm}
\includegraphics[width=2.1in]{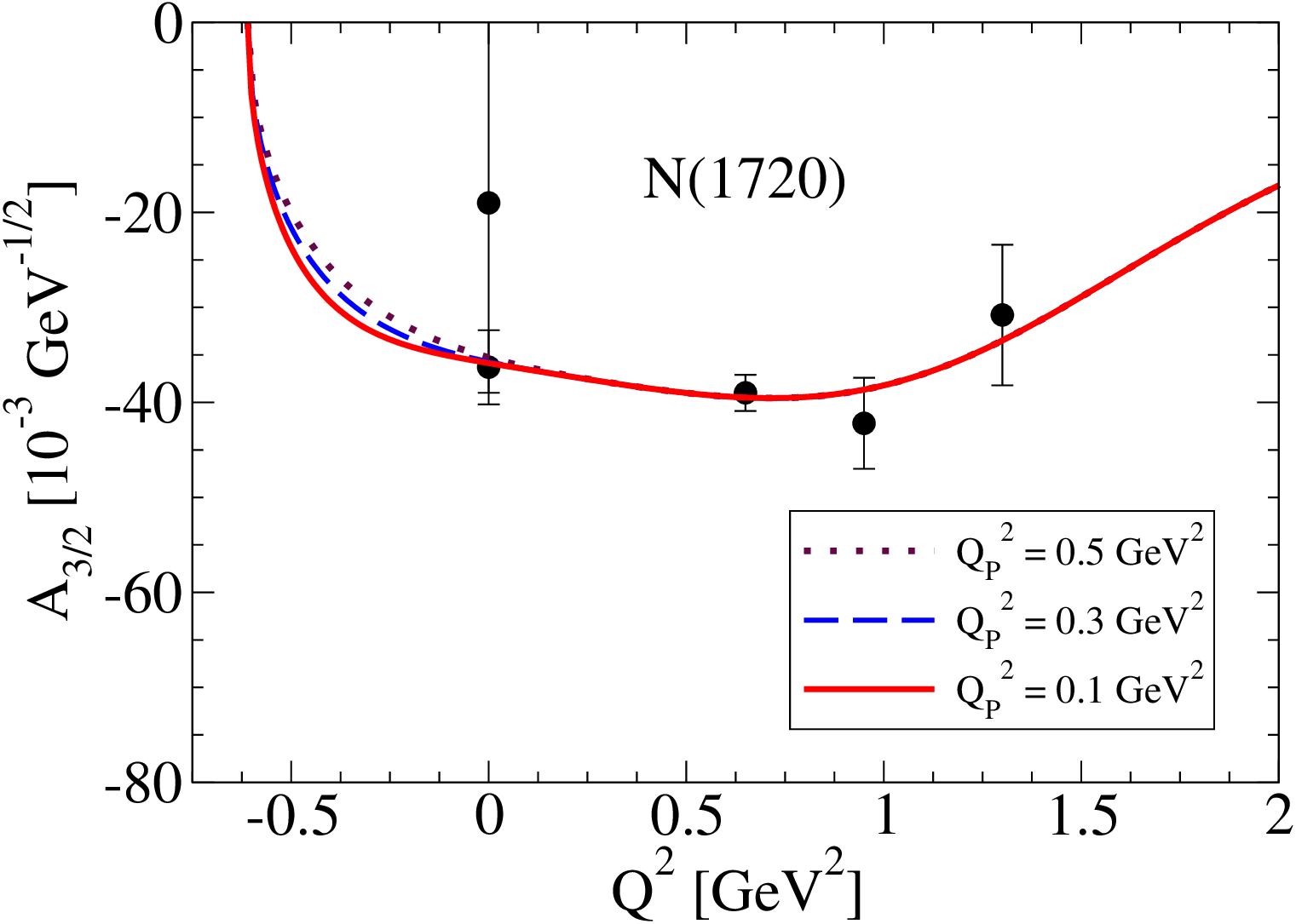}  \hspace{.3cm}
\includegraphics[width=2.1in]{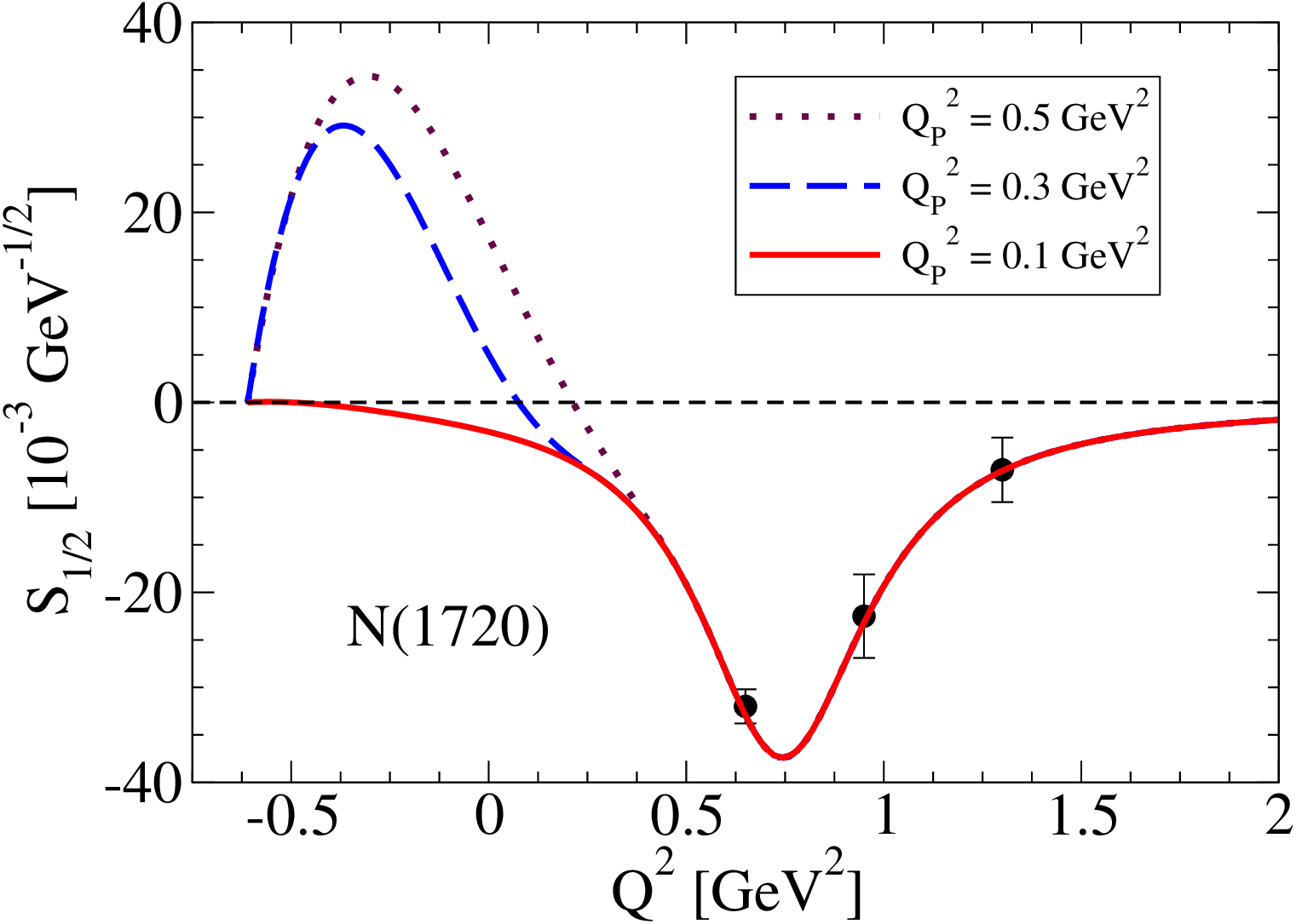} }}
\caption{\footnotesize{$\gamma^\ast N \to N^\ast
\left(\frac{3}{2}^+ \right)$ 
transition amplitudes. 
$N^\ast=$ $\Delta(1232)$, $N(1720)$.
Data for $\Delta(1232)$ from Refs.~\cite{PDG14,CLAS09,Stave08,Sparveris13,Blomberg16a}.
Data for $N(1720)$ from Refs.~\cite{PDG14,Mokeev16b}.
For the $N(1720)$ we include also the result at $Q^2=0$
from Refs.~\cite{Dugger09} used in the parametrization.}}
%\vspace{-1cm}
\label{fig-threehalfp}
\end{figure*}

Our results for the resonances 
$\frac{1}{2}^+$, $\frac{1}{2}^-$, $\frac{3}{2}^+$ and  $\frac{3}{2}^-$
are presented in Figs.~\ref{fig-onehalfp},  \ref{fig-onehalfm},
\ref{fig-threehalfp} and \ref{fig-threehalfm}, respectively.
The tables with the coefficients associated 
to each parametrization are presented in Appendix~\ref{appTables}.
In all cases, we include the original fit (solid dark line) 
in the $Q^2 > 0$ region,
although the line is not always visible due to the overlap of lines.

The states $\Delta(1232)\frac{3}{2}^+$, $N(1440)\frac{1}{2}^+$, 
$N(1520)\frac{3}{2}^-$ and  $N(1535)\frac{1}{2}^-$
are discussed first.
Later, we analyze the remaining cases.

\subsection{$N(1440)\frac{1}{2}^+$}

Our results for $N(1440)\frac{1}{2}^+$ are
at the top in Fig.~\ref{fig-onehalfp}.

The $N(1440)\frac{1}{2}^+$ state is interesting, because 
there is no direct relation between the two amplitudes.
The effect of the pseudothreshold 
is restricted to the leading-order dependence 
of the amplitudes on $|{\bf q}|$ near the pseudothreshold,
presented in Eqs.~(\ref{eqP11-1}).
Nevertheless, there are still interesting features 
in the different extension for the timelike region 
depending on the value of $Q^2_P$.

From the results for the amplitude $A_{1/2}$ 
we can conclude that the extrapolations for $Q^2 < Q_P^2$
are very sensitive to the value of $Q_P^2$.
This happens because the original parametrization has 
a large derivative near $Q^2=0$ in order to describe the $Q^2=0$ data. 
All the extrapolations attempt to provide a smooth transition 
between the result at $Q^2=0$ and the result at the pseudothreshold ($A_{1/2}=0$).
Note, however, that the parametrization with $Q_P^2=0.1$ GeV$^2$
has a strong deflection near $Q^2=0$ inducing a significant 
increment of the magnitude of the amplitude followed by 
a fast reduction to zero at the pseudothreshold.
The extension associated with  $Q_P^2=0.3$ GeV$^2$ 
is the one with the best balance between the description 
of the data and a smoother transition to the pseudothreshold,
and it is also close to the  $Q^2=0$ data point.
The $A_{1/2}$ amplitude provides a good example 
about the sensitivity of the parametrizations to the low-$Q^2$ data.

As for the amplitude $S_{1/2}$ all the parametrizations 
are consistent with the data, except for the 
lowest $Q^2$ point from MAMI 
(Mainz Microtron)
($Q^2=0.1$ GeV$^2$)~\cite{Stajner17}.

Only new data below $Q^2=0.3$ GeV$^2$ can help 
to determine the shape of $A_{1/2}$ and  $S_{1/2}$ 
near $Q^2=0$, and to select which analytic extension is better.

\subsection{$N(1535)\frac{1}{2}^-$}

Our results for the $N(1535)\frac{1}{2}^-$ are 
at the top in Fig.~\ref{fig-onehalfm}.
We discuss first the results represented by the thick lines.

From the graphs for $A_{1/2}$ we can conclude that 
all the extensions are equivalent.
The results for $S_{1/2}$, however, are very different.
The strong dependence of the analytic extensions on the point $Q_P^2$
is clear from the graph.
This result shows how relevant the  
pseudothreshold condition (\ref{eqS11-1}) is,
since the extrapolation demands a positive derivative 
for $S_{1/2}$ near the pseudothreshold.
It is worth mentioning that similar shapes 
can be obtained when we use 
phenomenological motivated parametrizations 
similar to the ones from MAID 
(combination of polynomials and exponentials)~\cite{Siegert1}.

The exact point where the function $S_{1/2}$ starts 
to approach zero when $Q^2$ decreases, 
still in the spacelike region ($Q^2> 0$),
cannot be determined by the data, 
since there are no $S_{1/2}$ data below $Q^2=0.3$ GeV$^2$.
Our extrapolation for the timelike region 
provides then an excellent example of how important 
measurements are below $Q^2=0.3$ GeV$^2$.

Our results for $N(1535)\frac{1}{2}^-$ 
provide another example of the impact of 
the pseudothreshold in the parametrizations of the data.
Based on the available $S_{1/2}$ data, 
we cannot distinguish between the three analytic extensions for $Q^2 < 0$.
Only future and accurate data can decide 
which extension for the $Q^2 < 0.5$ GeV$^2$ region is better.

In the present case it can be interesting to discuss 
the constraints imposed in 
the parametrizations of $A_{1/2}$ and $S_{1/2}$.
We recall that we chose the situation 
where the parametrization of 
$A_{1/2}$ is fixed by the continuity of the amplitude 
and the first three derivatives of the amplitude at $Q^2=Q_P^2$,
while for $S_{1/2}$ we demanded only the continuity of the 
amplitude and the first two derivatives.
We impose then stronger conditions for $A_{1/2}$, 
since as mentioned the function $A_{1/2}$ is constrained 
more accurately by the data,
since the error bars are smaller and we have an estimate of $A_{1/2}$ at $Q^2=0$.
As a consequence of those constraints, all extensions of $A_{1/2}$
are described by smooth functions, almost undistinguished 
between them (overlap of thick lines).
All those extensions are consistent with a very small 
third derivative of $A_{1/2}$ near $Q^2=0$.

From the graph for $A_{1/2}$,
one can notice, however, that the data point at $Q^2=0$ 
has a very large errorbar. 
In the present case, this happens because the data selected by the PDG
have a large dispersion, and the PDG considers a very large window of variation.
Since  $A_{1/2}(0)$ is poorly determined, one can question 
if the use of the amplitude  $A_{1/2}$  as a reference is, in fact, a good choice
and if we should not consider the hypothesis of 
using the amplitude $S_{1/2}$ as a reference, 
although based on data with larger errorbars.
If we use this alternative procedure, 
we are replacing a parametrization derived from the 
condition that the third derivative of $A_{1/2}$ 
is small ($A_{1/2}^{(3)} \approx 0$),
by a parametrization that assumes that it is the function $S_{1/2}$
that is smooth, with coefficients determined by 
the first three derivatives of $S_{1/2}$.
In that case we used Eq.~(\ref{eqS11-1}) to determine 
the shape of $A_{1/2}$ near the pseudothreshold
based on the shape of $S_{1/2}$. 

In order to test the impact of this alternative method 
to the pseudothreshold constraints,
we recalculate the parameters of the amplitudes 
$A_{1/2}$ and $S_{1/2}$ based on the previous hypothesis:
Fix $S_{1/2}$ by the first three derivatives 
and $A_{1/2}$ by Eq.~(\ref{eqS11-1}) and the first two derivatives,
also for the points $Q_P^2=0.1$, 0.3, and 0.5 GeV$^2$.
The new estimates are represented in the graphs by the thin 
lines with the same convention.
It is clear from the graphs that $A_{1/2}$ is not described 
by smooth functions near $Q^2=0$.
The results for $A_{1/2}$ are now described 
by functions consistent with a large $A_{1/2}^{(3)}$.

As for the amplitude $S_{1/2}$, it is now described 
by very smooth functions of $Q^2$ in the timelike region.
All the extensions of $S_{1/2}$ are now very similar. 

We emphasize that both descriptions 
(thin lines and thick lines) are consistent with the available data, 
where one has a gap in the region $Q^2=0$--0.3 GeV$^2$,
and the large error for $A_{1/2}(0)$. 
One concludes, then that only new data 
for $S_{1/2}$ or more accurate result for $A_{1/2}(0)$ can decide which 
description of the low-$Q^2$ data is the best.

% FIGURE 4

\subsection{$\Delta(1232)\frac{3}{2}^+$}

Our results for the $\Delta(1232)\frac{3}{2}^+$ are 
at the top in Fig.~\ref{fig-threehalfp}.

From the graphs we can conclude that 
the parametrization characterized by $Q_P^2=0.3$ GeV$^2$
(dashed line) is the one that better describes the data,
more specifically, the $Q^2 < 0.3$ GeV$^2$ data.
There are two main explanations for this result.
A possible explanation is the fact that 
the parametrizations from Ref.~\cite{Jlab-fits} are based 
on simple rational functions of $Q= \sqrt{Q^2}$ 
and, therefore, do not try to describe in detail the low-$Q^2$ region, 
since they are more focused on the  $Q^2=0.5$--5 GeV$^2$ region.
Another explanation is that the photon point ($Q^2=0$) 
is in the present case very close to the pseudothreshold 
$Q^2=- (M_R-M_N)^2  \simeq -0.09$ GeV$^2$.
The consequence of the closeness between those two points 
is that the pseudothreshold constraints (\ref{eqP33-1}),
and $A_{1/2}, A_{3/2} \propto |{\bf q}|$ demand a sharp variation 
between the results near $Q^2=0$ and the pseudothreshold, 
where all the amplitudes vanish, although with different rates.
%

% FIGURE 4

\begin{figure*}[t]
%\vspace{0.5cm}
\centerline{\vspace{0.5cm}  }
\centerline{
\mbox{
\includegraphics[width=2.1in]{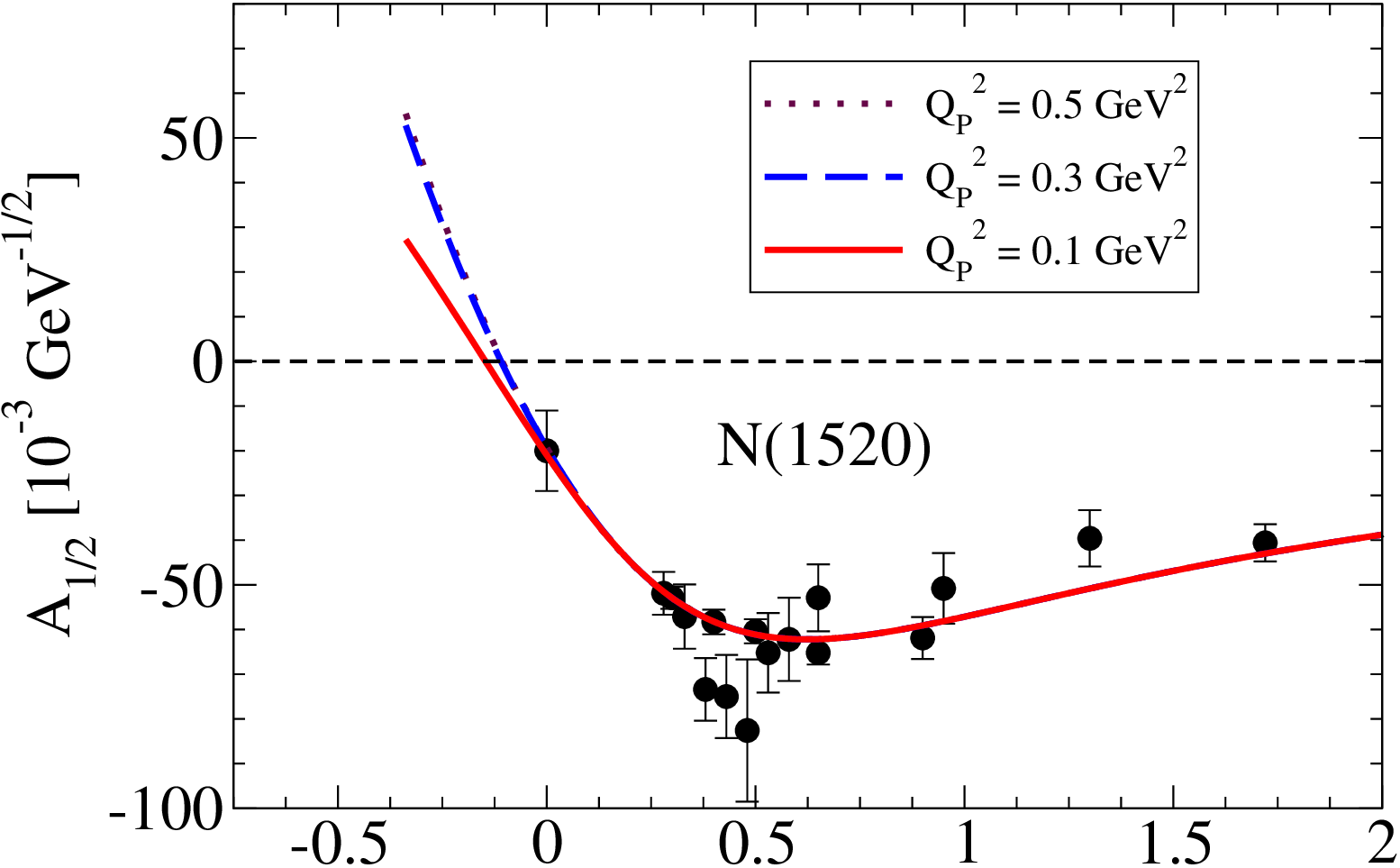} \hspace{.3cm}
\includegraphics[width=2.1in]{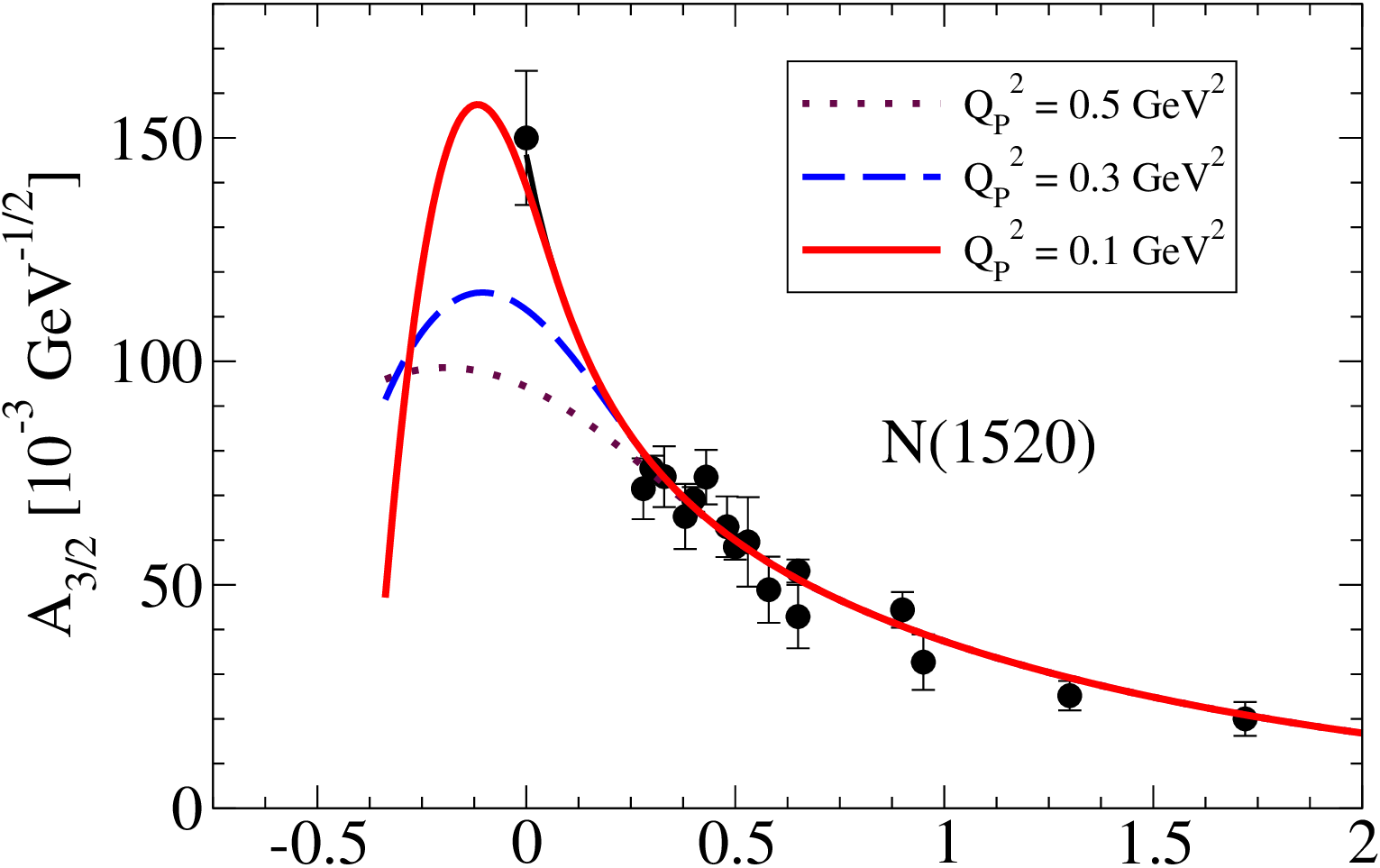}  \hspace{.3cm}
\includegraphics[width=2.1in]{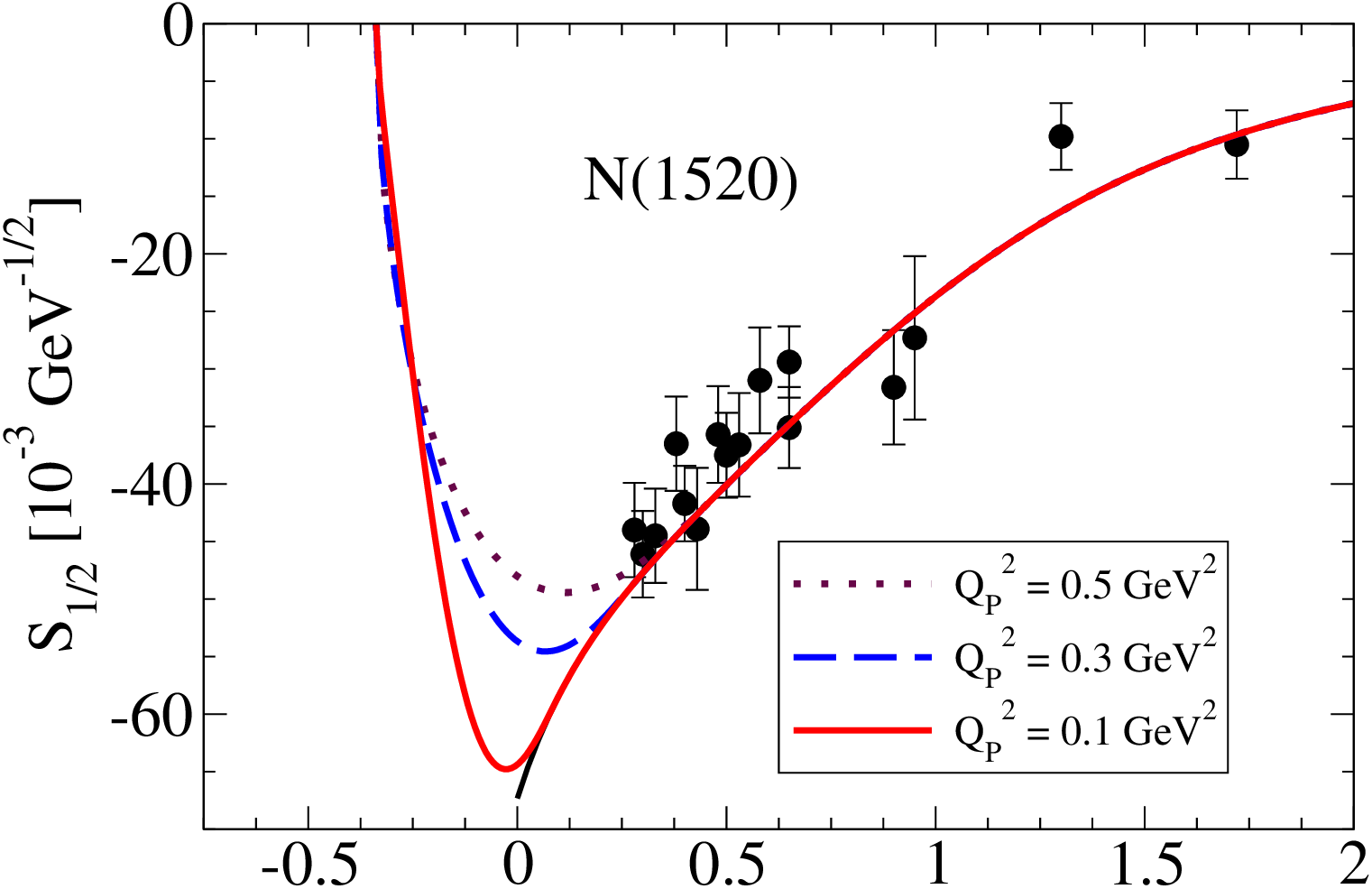} }}
\vspace{.7cm}
\centerline{
\mbox{
\includegraphics[width=2.1in]{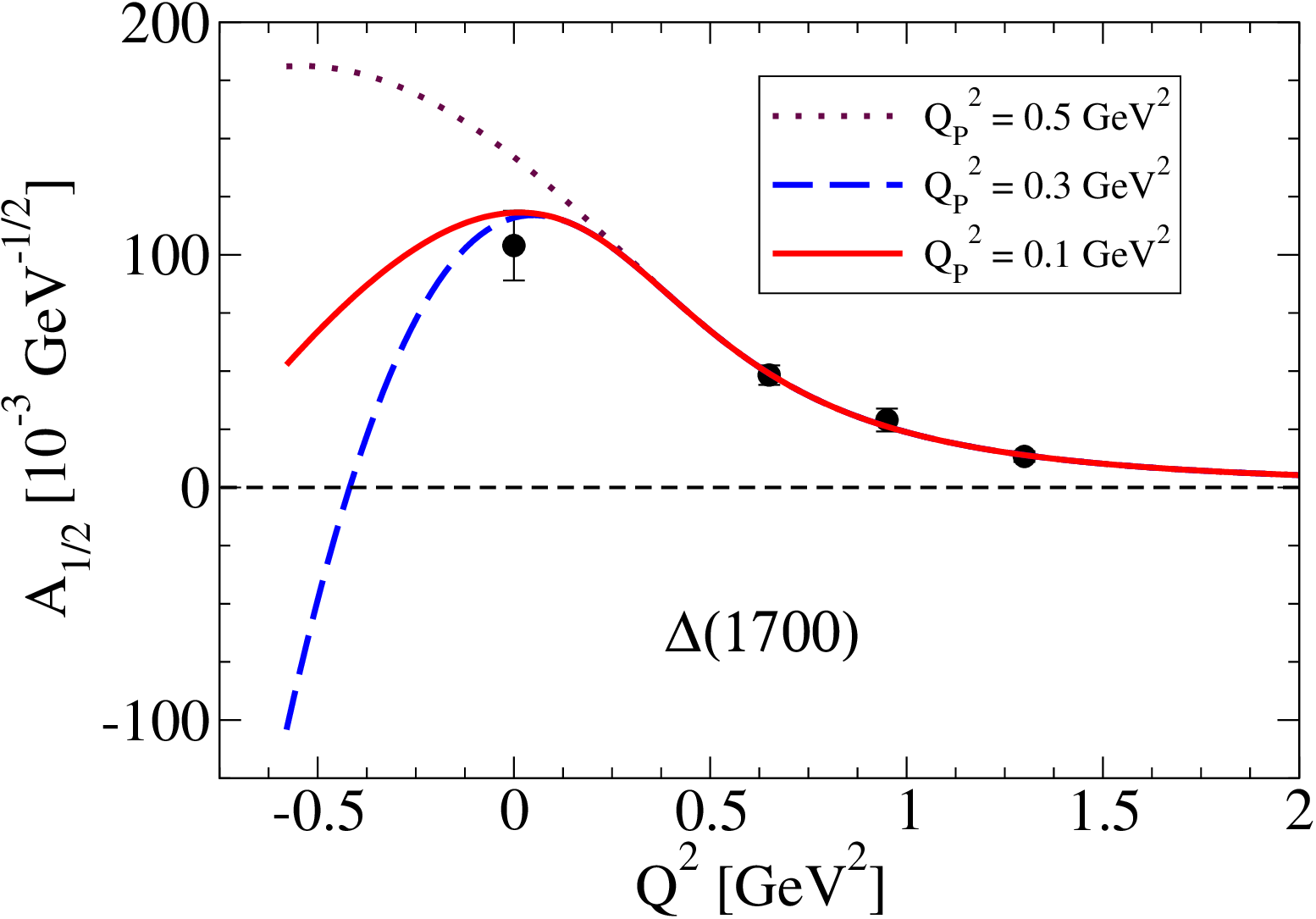} \hspace{.3cm}
\includegraphics[width=2.1in]{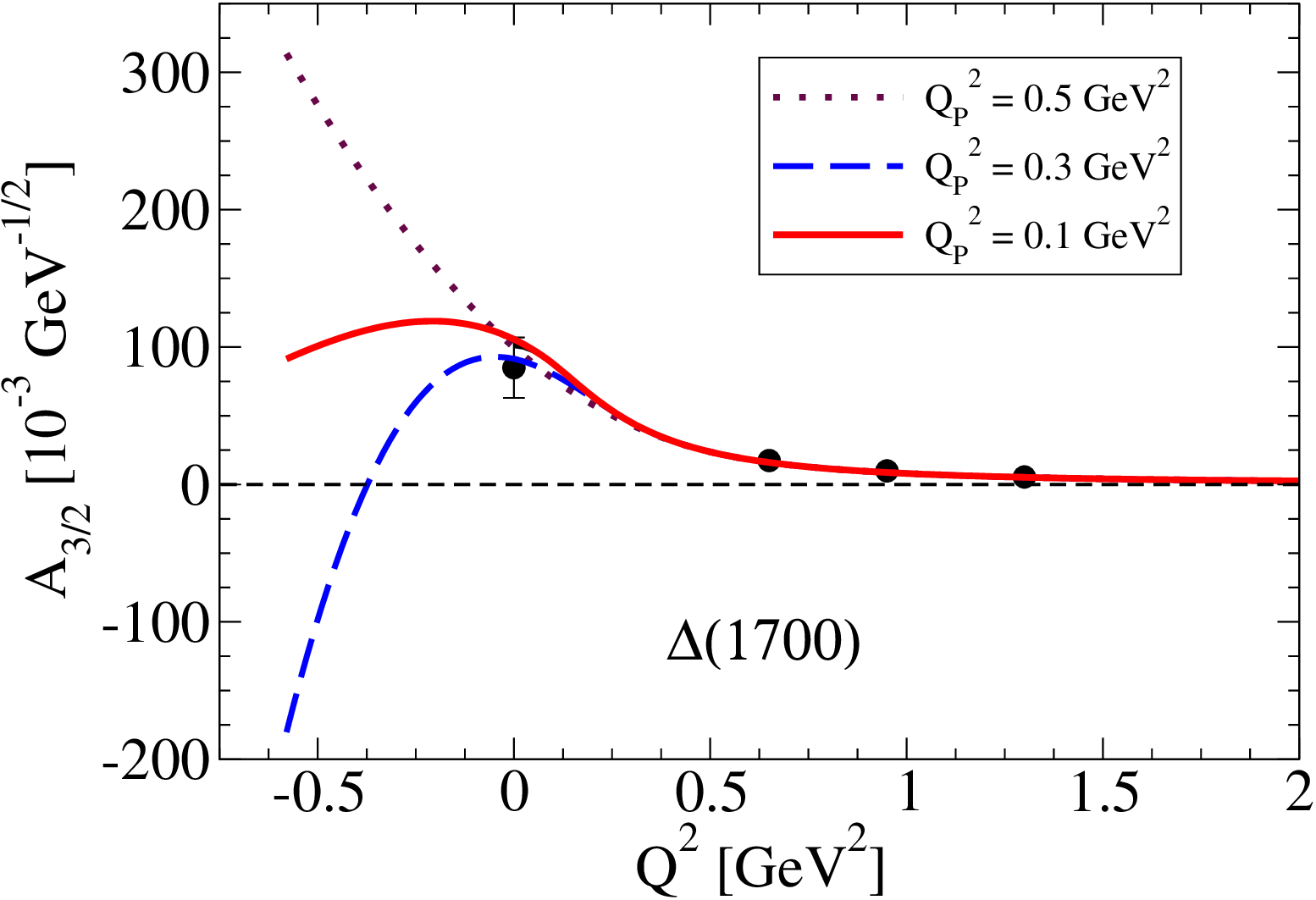}  \hspace{.3cm}
\includegraphics[width=2.1in]{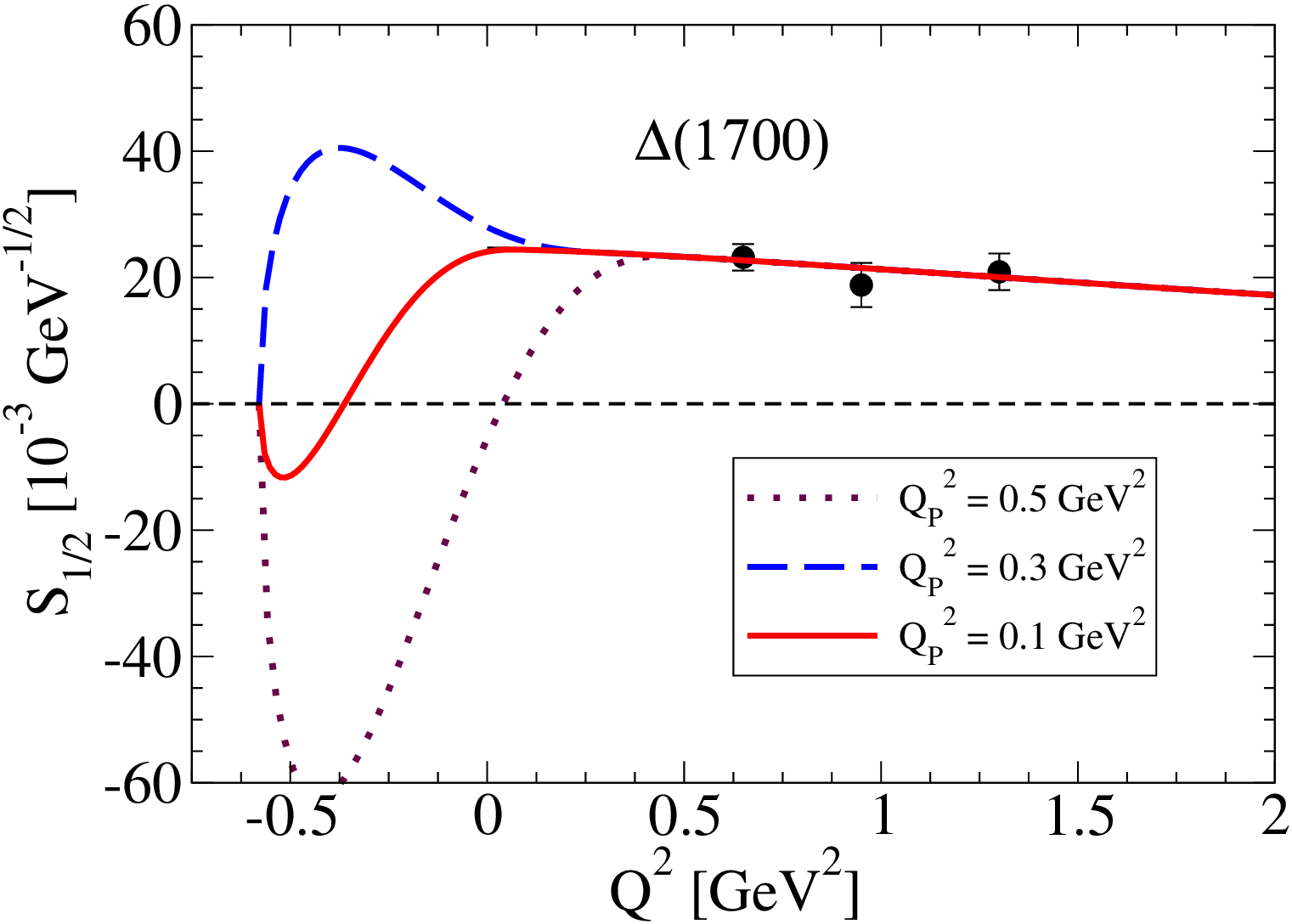} }}
\caption{\footnotesize{$\gamma^\ast N \to 
N^\ast \left( \frac{3}{2}^- \right)$  transition amplitudes.
$N^\ast=$ $N(1520)$, $\Delta(1700)$.
Data for $N(1520)$ from Refs.~\cite{PDG14,CLAS09,CLAS12,Mokeev16}.
Data for $\Delta(1700)$ from Refs.~\cite{PDG14,Mokeev14}.}}
%\vspace{-1cm}
\label{fig-threehalfm}
\end{figure*}

The sharp variation near the pseudothreshold 
can be understood in simple terms, 
noting that, using the dominance of the magnetic form 
factor ($G_M$) over the electric form factor 
($G_E$), we can write~\cite{Jones73,NDelta}
\ba
A_{1/2} = - {\cal B} G_M, 
\hspace{.3cm}
A_{3/2} = - \sqrt{3} {\cal B} G_M, 
\ea
where 
\ba
{\cal B}  = \frac{1}{2}\frac{M_R}{M_N} 
\sqrt{\frac{\pi \alpha}{M M_R K}} \frac{|{\bf q}|}{\sqrt{1 + \tau}} 
\propto |{\bf q}|.
\ea
In the region $-(M_R-M_N)^2 \le Q^2 \le 0$ 
one can approximate $\sqrt{1 + \tau} \simeq 1$ 
with an accuracy better than 1\%.
In these conditions $A_{1/2}$ and  $A_{3/2}$ 
are determined exclusively by $G_M$.

Assuming that $G_M$ is a smooth function of $Q^2$ 
in the interval $[ - (M_R -M_N)^2, 0]$, we can write
\ba
& &
G_M (Q^2)  \simeq  G_M(0) + G_M^\prime (0) Q^2 \nonumber \\
       &  & \simeq  
 [G_M(0) - G_M^\prime (0) (M_R -M_N)^2 ] + G_M^\prime (0) \frac{M_R}{M_N}  
|{\bf q}|^2, \nonumber \\
\ea
where the use the expansion
$Q^2 \simeq -(M_R -M_N)^2 + \frac{M_R}{M_N}  |{\bf q}|^2$,
for small  $|{\bf q}|$ based on Eq.~(\ref{eqQ2a}).

A simple estimate with the MAID 2007 parametrization~\cite{Drechsel2007} 
gives $G_M (Q^2) \simeq 3.75 - 10.17 \frac{|{\bf q}|^2}{M_N^2}$.
Based on the previous result, one obtains
\ba
A_{1/2} \simeq \frac{A_{3/2}}{\sqrt{3}} 
= - b \left(  3.75  -  10.17 \frac{|{\bf q}|^2}{M_N^2}
\right)  |{\bf q}|,
\ea
where $b= 0.0908$ GeV$^{-3/2}$.
The leading-order term is then corrected 
by a term in $|{\bf q}|^3$ ensuring a smooth
transition to the $Q^2 > 0$ region.

The study of the  pseudothreshold constraints 
on the $\gamma^\ast N \to \Delta(1232)$ transition 
can also be performed using the form factor representation, 
based on Siegert's theorem (\ref{eqD33}).
From the theoretical point of view, the electric and Coulomb form factor data 
at low $Q^2$ can be described by a combination 
of valence quark and pion cloud effects~\cite{LatticeD},
where both contributions are compatible 
with the relation (\ref{eqD33})~\cite{DeltaQuad,RSM-Siegert,GlobalP}.

The data presented  in Fig.~\ref{fig-threehalfp}
deserve some discussion.
Contrary to most of the resonances 
there are for the $\Delta(1232)$ finite $Q^2$ data below $Q^2=0.3$ GeV$^2$.
The database from Ref.~\cite{MokeevDatabase} 
includes data from CLAS~\cite{CLAS09} and low-$Q^2$ data  
from different sources such as 
MAMI~\cite{Stave08}, MIT-Bates~\cite{MIT_data,Sparveris13}, 
and data at $Q^2=0$ from the PDG~\cite{PDG14}.
In the present study, however, we replaced the data for $Q^2 < 0.15$ GeV$^2$
by more recent estimates of the data.
Concerning the data for $Q^2=0$ for $A_{1/2}$ and $A_{3/2}$,
we use the data associated to $G_M(0)$ 
and the ratio $R_{EM}= - \frac{G_E(0)}{G_M(0)}$,
also from the PDG~\cite{PDG14}.
This procedure is justified by the difference 
of results for the form factors obtained when we use
the helicity amplitudes~\cite{Pascalutsa07b}.

As for the $Q^2< 0.15$ GeV$^2$ data,
we replace the results from MAMI and MIT-Bates 
($Q^2= 0.06$ and 0.127 GeV$^2$)~\cite{Stave08,MIT_data,Sparveris13}
by the recent results from JLab/Hall A
($Q^2= 0.09$ and 0.13 GeV$^2$)~\cite{Blomberg16a}.
This procedure is motivated by the conclusion 
that there are errors in the previous 
analysis which lead to an overestimation 
of the results for $G_E$ and $G_C$,
as discussed in Refs.~\cite{Blomberg16a,RSM-Siegert}.
The data for $A_{1/2}$, $A_{3/2}$ and $S_{1/2}$ 
presented here are converted from the results 
for the form factors presented in Refs.~\cite{RSM-Siegert,GlobalP}. 
For the conversion, we used the MAID 2007 parametrization~\cite{Drechsel2007} 
for the magnetic form factor, since it is simple and accurate 
in the region of study.

\subsection{$N(1520)\frac{3}{2}^-$}

Our results for the $N(1520)\frac{3}{2}^-$ are 
at the top in Fig.~\ref{fig-threehalfm}.

From the graphs we conclude that the 
amplitudes $A_{3/2}$ and $S_{1/2}$ are very sensitive 
to $Q_P^2$, while the amplitude 
$A_{1/2}$ is almost independent of $Q_P^2$ in the $Q^2 > 0$ region. 
From all the analytic extensions only the parametrizations
with $Q_P^2= 0.1$ GeV$^2$ are consistent with the data for $A_{3/2}(0)$.
As for the amplitude $S_{1/2}$, all the extensions are 
valid, since there are no data for $Q^2 < 0.3$ GeV$^2$.

Overall, we can conclude that the parametrizations with  
$Q_P^2= 0.1$ GeV$^2$ provide a smooth extrapolation 
to the timelike region and are consistent with the available data.

\subsection{Other resonances}

The remaining resonances 
show that some amplitudes are very sensitive 
to the threshold of the extrapolation 
(value of $Q_P^2)$.
This result is also a consequence 
of the limited data used in the calibrations 
(between three and five points)
combined with the nonexistent data below $Q^2=0.5$ GeV$^2$,
except for $Q^2=0$.

The exception is the state $N(1710)\frac{1}{2}^+$.
Those results are a consequence of the lack 
of constraints, apart from the relations (\ref{eqP11-1}),
and also to the more significant distance 
between the $Q^2=0$ data and the pseudothreshold.
This distance is about $ 0.6$ GeV$^2$,
much larger than the $N(1440)$ case  ($\approx 0.25$ GeV$^2$).
In the case of the $N(1710)$, there is more room for the amplitudes to fall down 
to the pseudothreshold.

Concerning the resonance $\Delta(1620)\frac{1}{2}^-$, 
a note about the parametrizations from Ref.~\cite{Jlab-fits} is in order.
The original parametrization has the form $S_{1/2} \propto 1/Q^3$,
meaning that it diverges  in the limit $Q^2 \to 0$.
We modified the parametrization to $S_{1/2} \propto 1/(\Lambda^3 + Q^3)$ 
using a finite value to $\Lambda$ in order 
to derive our analytic extension to $Q^2 < 0$.
We use, in particular $\Lambda^3 = 0.1$ GeV$^3$.
With this modification, we obtain 
a parametrization close to the original data 
and avoid the divergence of $S_{1/2}$. 
Nevertheless, we obtain very large values for  the magnitude of 
$S_{1/2}$ near $Q^2=0$, and below that point.

\section{Outlook and conclusions}
\label{secConclusions}

In the present work we analyze the impact of 
the pseudothreshold constraints in the empirical 
parametrizations of the 
$\gamma^\ast N \to N^\ast$ transition amplitudes.
The formalism proposed can be applied to any 
analytic parametrizations of the transition amplitudes
in the spacelike region.
To exemplify the potential of the method,
we restrict the applications to the JLab parametrizations
from Ref.~\cite{Jlab-fits}, since they cover 
the region $Q^2=0$--5 GeV$^2$, and provide 
a good description of the overall data, in general,  
and the large $Q^2$ from CLAS/JLab in particular.

The sensibility of the parametrizations 
is determined looking to the point $Q_P^2$, 
where the parametrizations are modified in order 
to be consistent with the pseudothreshold conditions, 
demanded by the structure of the transition current.
Motivated by the $Q^2$ distribution of the measured data
which have in general a gap in the region $Q^2=0$--0.3 GeV$^2$, 
we derived analytic continuations for 
the region between the pseudothreshold 
$Q^2= -(M_R-M_N)^2$ up to the point $Q_P^2$,
for $Q_P^2= 0.1$, 0.3 and 0.5 GeV$^2$.
  
For resonances characterized by a small number of data points 
(three to five),
our results are not conclusive, either by the lack 
of low-$Q^2$ data ($Q^2 < 0.5$ GeV$^2$), or because 
the pseudothreshold is far away from the photon point.

As for the more well-known resonances:
$\Delta(1232)\frac{3}{2}^+$, $N(1440)\frac{1}{2}^+$, 
$N(1520)\frac{3}{2}^-$, and $N(1535)\frac{1}{2}^-$, 
the results are very relevant for different reasons.

Our results for the $\Delta(1232)$ 
show  conclusively that the constraints at the pseudothreshold 
cannot be ignored below 0.3 GeV$^2$.
This result is also a consequence of 
the closeness between the pseudothreshold 
($\approx -0.09$ GeV$^2$) and the photon point ($Q^2=0$).

For the $N(1440)$, we conclude that the
parametrizations are almost insensitive 
to the pseudothreshold conditions 
except for the region $Q^2 < 0.3$ GeV$^2$.
More low-$Q^2$ data are necessary to determine the 
correct shape of $A_{1/2}$ below $Q^2 < 0.3$ GeV$^2$.

The results for the $N(1520)$ manifest a significant  
dependence of the amplitudes $A_{3/2}$ and $S_{1/2}$ 
on the value of $Q_P^2$. 
Only the extension with $Q_P^2=0.1$ GeV$^2$ 
is consistent with the available data,
meaning that the pseudothreshold constraints 
are relevant only for $Q^2 < 0.1$ GeV$^2$.

Finally, for the $N(1535)$, we obtain several
analytic extensions to the timelike region which are
compatible with the available data.
One concludes that accurate measurements 
of $A_{1/2}$ at the photon point and 
new measurements of the longitudinal amplitude
$S_{1/2}$ are fundamental to determine 
the shape of the two amplitudes below $Q^2=0.3$ GeV$^2$.

Overall, we conclude that the impact 
of the pseudothreshold conditions 
can be observed definitely in the case of the $\Delta(1232)$
near $Q^2= 0.3$ GeV$^2$.
A soft transition to the pseudothreshold limit 
can also be observed in the $N(1520)$.
In the remaining cases, 
there are parametrizations compatible with 
the pseudothreshold conditions, 
but the upcoming data from 
the JLab 12 GeV upgrade will be very important to pin down 
the shape of the transition amplitudes below $Q^2=0.3$ GeV$^2$.

%\clearpage

\begin{acknowledgments}
%\vspace{-.4cm}
G.~R.~was supported by the Funda\c{c}\~ao de Amparo \`a
Pesquisa do Estado de S\~ao Paulo (FAPESP):
Project No.~2017/02684-5, Grant No.~2017/17020-BCO-JP.
\end{acknowledgments}

\appendix

\section{Determination of the coefficients $\alpha_l$}
\label{appendix-alpha}

We consider here the amplitude $A$ from Eq.~(\ref{eqPowerQT2}) with $n=0$.
The other cases can be extrapolated using 
the functions $A/\tilde q$ or $A/\tilde q^2$.

The coefficients $\alpha_l$ 
of the function $A$ from Eq.~(\ref{eqPowerQT2}) 
can be determined using the 
continuity of the functions $A$, $\frac{d A}{d \tilde q^2}$,
and $\frac{d^2 A}{d \tilde q^4}$ at the point $Q^2=Q_P^2$.
For convenience, we represent the 
moments of the expansion (\ref{eqPowerQT2}) by
\ba
\tilde A^{(k)} = \frac{d^k A}{d \tilde q^{2k}} (Q_P^2),
\ea
for $k=0,1,2,3$.
The conversion between the 
moments of the expansion in $Q^2$ ($A^{(k)}$) into the moments 
of the expansion in $\tilde q^2$ ($\tilde A^{(k)}$)
is presented in Appendix~\ref{appDerivatives}.

The continuity of $A$, $\frac{d A}{d \tilde q^2}$,
and $\frac{d^2 A}{d \tilde q^4}$
implies that
\ba
& &
\alpha_0 + \alpha_1 \tilde q_p^2 + 
 \alpha_2 \tilde q_p^4 +  \alpha_3 \tilde q_p^6 = \tilde A^{(0)}, 
\\
& &
\alpha_1 + 2\alpha_2 \tilde q_p^2 + 
 3 \alpha_3 \tilde q_p^4  = \tilde A^{(1)},
 \\
& &
2\alpha_2  + 
 6 \alpha_3 \tilde q_p^2  = \tilde A^{(2)},
\ea
where $\tilde q_p^2$ represent $\tilde q^2$
for $Q^2= Q_P^2$. 

From the previous equations, one obtains
\ba
& &
\alpha_2 = \frac{1}{2} \tilde A^{(2)} - 3 \alpha_3 \tilde q_p^2
\label{eqA2}\\
& &
\alpha_1 = \tilde A^{(1)} -  \tilde A^{(2)} \tilde q_p^2 +  
3 \alpha_3 \tilde q_p^4\\
& &
\alpha_0 = \tilde A^{(0)} -  \tilde A^{(1)} \tilde q_p^2 
+ \sfrac{1}{2} \tilde A^{(2)} \tilde q_p^4 -  \alpha_3 \tilde q_p^6.
\label{eqA0}
\ea
These equations can be used 
to calculate $\alpha_0$,  $\alpha_1$ and  $\alpha_2$ 
once the value of $\alpha_3$ is fixed.

In the present work we use Eqs.~(\ref{eqA2})--(\ref{eqA0})
to calculate the coefficients 
based on two different conditions:
\begin{enumerate}
\item
The coefficient $\alpha_3$ is determined 
using the continuity of the third derivative: 
\ba
6 \alpha_3 = \tilde A^{(3)};
\label{eqA3}
\ea
\item
the coefficient  $\alpha_0$ is determined by 
a pseudothreshold condition.
The coefficient $\alpha_3$ can then 
be determined by Eq.~(\ref{eqA0}) using 
\ba
\alpha_3 = (\tilde A_p - \alpha_0) \frac{1}{\tilde q_p^6},
\label{eq-alpha3}
\ea
where $\tilde A_p \equiv \tilde A^{(0)} -  \tilde A^{(1)} \tilde q_p^2 
+ \sfrac{1}{2} \tilde A^{(2)} \tilde q_p^4$.
\end{enumerate}

According to the discussion from Sec.~\ref{secAnalytic},
the first condition is used in case 1, and 
the second condition is used in case 2.

In general Eq.~(\ref{eqA3}) is used 
only for the amplitudes $A_{1/2}$ and $A_{3/2}$,
depending on the $J^P$ state.
The exception is the amplitude $S_{1/2}$ for the 
$\frac{1}{2}^+$, for which there is no particular 
correlation between amplitudes.

%\vspace{.5cm}

\setcounter{table}{0}
\renewcommand{\thetable}{B\arabic{table}}

\begin{table*}[b]
\begin{center}
\begin{tabular}{l l}
\hline
\hline 
%\vspace{.1cm}
 $n=0$ & \\[.1cm]
 $A = a_0 + a_1 \tilde q^2  + a_2 \tilde q^4 +   a_3 \tilde q^6$  & 
$\tilde A^{(0)} \equiv A$
\\[.1cm]
 $\frac{d A}{ d \tilde q^2} =
a_1 + 2  a_2 \tilde q^2 +   3 a_3 \tilde q^4$ & 
$ \tilde A^{(1)} \equiv 
\frac{d A}{ d \tilde q^2} =Z M_R^2 \frac{d A}{d Q^2}$ \\[.1cm]
 $\frac{d^2 A}{ d \tilde q^4} =
 2 a_2  +   6 a_3 \tilde q^2$ & $
\tilde A^{(2)} \equiv 
\frac{d^2 A}{ d \tilde q^4} =  
Z^2  M_R^4 \frac{d^2 A}{ d Q^4} - 2 Z^3 M_R^2  \frac{d A}{d Q^2}
 $ \\[.1cm]
 $\frac{d^3 A}{ d \tilde q^6} =   6 a_3 $ & 
$\tilde A^{(3)} \equiv 
\frac{d^3 A}{ d \tilde q^6} = 
Z^3 M_R^6  \frac{d^3 A}{d Q^6}  
- 6 Z^4 M_R^4 \frac{d^2 A}{d Q^4}
+ 12 Z^5 M_R^2 \frac{d A}{d Q^2}
$ \\[.1cm]
&  \\ 
$n=1$ & \\[.1cm]
  $ \frac{A}{\tilde q} = a_0 + a_1 \tilde q^2  + a_2 \tilde q^4 +   a_3 \tilde q^6$ 
&  $\tilde A^{(0)} \equiv \frac{A}{\tilde q}$\\[.1cm]
$\frac{d \; }{ d \tilde q^2} \left( \frac{A}{\tilde q} \right)=
a_1 + 2  a_2 \tilde q^2 +   3 a_3 \tilde q^4$  \spQ
& $\tilde A^{(1)} \equiv \frac{d}{d \tilde q^2} 
\left( \frac{A}{\tilde q} \right) =
 \frac{Z}{\tilde q}   M_R^2 \frac{d A}{d Q^2}  -  
\frac{1}{2 \tilde q^3} A $  \\[.1cm]
$\frac{d^2 \; }{ d \tilde q^4} \left( \frac{A}{\tilde q} \right)=
2  a_2  +   6 a_3 \tilde q^2$ 
& $\tilde A^{(2)} \equiv
\frac{d^2 }{ d \tilde q^4} \left( \frac{A}{\tilde q} \right) =
\frac{Z^2}{\tilde q}  M_R^4 \frac{d^2 A}{d Q^4} 
- d_{11} M_R^2 \frac{d A}{d Q^2}
+ \frac{3}{4 \tilde q^5} A  
$ \\[.1cm]
 $\frac{d^3 }{ d \tilde q^6} \left( \frac{A}{\tilde q} \right)  
=   6 a_3 $ &   
$\tilde A^{(3)} \equiv \frac{d^3 }{ d \tilde q^6} 
\left( \frac{A}{\tilde q} \right) =
\frac{Z^3}{\tilde q}  M_R^6 \frac{d^3 A}{d Q^6} 
- d_{12} M_R^4 \frac{d^2 A}{d Q^4}  + d_{13}  
M_R^2 \frac{d A}{d Q^2} - \frac{15}{8} \frac{1}{\tilde q^7} A
$  \\[.25cm]
& $d_{11}= 2\frac{Z^3}{\tilde q} +  \frac{Z}{\tilde q^3}$, 
\hspace{.15cm}
$d_{12}=  6 \frac{Z^4}{\tilde q} + \frac{3}{2}\frac{Z^2}{\tilde q^3} $,
\hspace{.15cm}
$d_{13}=  12 \frac{Z^5}{\tilde q} +  3 \frac{Z^3}{\tilde q^3}
+ \frac{9}{4}\frac{Z}{\tilde q^5} $
\\[.1cm]
& \\ 
$n=2$ & \\[.1cm]
  $\frac{A}{\tilde q^2} = 
a_0 + a_1 \tilde q^2  + a_2 \tilde q^4 +   a_3 \tilde q^6$ 
& $\tilde A^{(0)} \equiv \frac{A}{\tilde q^2}$
\\
 $\frac{d \; }{ d \tilde q^2} \left( \frac{A}{\tilde q^2} \right)=
a_1 + 2  a_2 \tilde q^2 +   3 a_3 \tilde q^4$
& $ \tilde A^{(1)} \equiv
\frac{d}{d \tilde q^2} \left( \frac{A}{\tilde q^2} \right) =
 \frac{Z}{\tilde q^2}  M_R^2  \frac{d A}{d Q^2}
 -  \frac{1}{\tilde q^4} A  $ \\[.1cm]
 $\frac{d^2 \; }{ d \tilde q^4} \left( \frac{A}{\tilde q^2} \right)=
2  a_2  +   6 a_3 \tilde q^2$
& $\tilde A^{(2)} \equiv
 \frac{d^2}{d \tilde q^4} \left( \frac{A}{\tilde q^2} \right) =
 \frac{Z^2}{\tilde q^2}  M_R^4  \frac{d^2 A}{d Q^4}
- 2 d_{21} 
%\left( \frac{Z^3}{\tilde q^2} + \frac{Z}{\tilde q^4} \right) 
M_R^2  \frac{d A}{d Q^2} + 
   \frac{2}{\tilde q^6} A$ \\[.1cm]
 $\frac{d^3 }{ d \tilde q^6} \left( \frac{A}{\tilde q^2} \right)  
=   6 a_3 $ &   
$\tilde A^{(3)} \equiv
\frac{d^3 }{ d \tilde q^6} \left( \frac{A}{\tilde q^2} \right) =
\frac{Z}{\tilde q^2} M_R^6 \frac{d^3 A}{d Q^6} 
- 3 d_{21} Z M_R^4  \frac{d^2 A}{d Q^4} +
6  d_{22}  M_R^2  \frac{d A}{d Q^2} - \frac{6}{\tilde q^8} A$ \\[.25cm]
& 
$d_{21} = \frac{Z^3}{\tilde q^2} +  \frac{Z}{\tilde q^4} $, 
\hspace{.15cm}
$d_{22} = 2\frac{Z^5}{\tilde q^2} +  \frac{Z^3}{\tilde q^4} + 
\frac{Z}{\tilde q^6}$ \\
\hline
\hline
\end{tabular}
\end{center}
\caption{\footnotesize
Relations between derivatives in $\tilde q^2$ 
and in $Q^2$ for the different kind of amplitudes
($A$ with $n=0,1,2$).
To compact the notation, we use the coefficients $d_{ij}$.}
\label{tabDerivatives}
\end{table*}

\section{Derivatives of amplitudes in $\tilde q^2$}
\label{appDerivatives}

% Table B1

In the present appendix, we derive the expressions
necessary to calculate the moments $\tilde A^{(k)}$
of the expansion (\ref{eqPowerQT2}). 
The coefficients $\alpha_l$ associated with the 
amplitudes $A_{1/2}$, $A_{3/2}$ and $S_{1/2}$
can then be determined using the values  $\tilde A^{(k)}$
and the relations derived in Appendix~\ref{appendix-alpha}.

For the purpose of the discussion, 
we consider a generic amplitude
\ba
A (\tilde q^2) =
\tilde q^n 
(a_0 + a_1 \tilde q^2 + a_2 \tilde q^4 + a_3 \tilde q^6), 
\ea
where $n=0,1,2$ and
$a_l$  ($l=0,1,2,3$) are real numbers.

Depending of the helicity amplitude under discussion, 
we need to calculate the following derivatives
\ba
\frac{d  A }{d  \tilde q^2} , 
\hspace{.2cm}
\frac{d }{d \tilde q^2} \left( \frac{A}{\tilde q} \right),
\hspace{.2cm}
\frac{d }{d \tilde q^2} \left( \frac{A}{\tilde q^2} \right), ...
\ea
For convenience, we use
\ba
\frac{d Q^2}{d \tilde q^2} = M_R^2 Z,
\ea
where 
\ba
Z =  \frac{ 2 M_R^2}{M_R^2 + M_N^2 + Q^2}.
\ea

The results for the different derivatives in $\tilde q^2$ 
for the cases $n=0,1,2$ are presented in Table~\ref{tabDerivatives}.

% TABLE B1

\newpage

%\clearpage

\setcounter{table}{0}
\renewcommand{\thetable}{C\arabic{table}}

\section{Coefficients associated with the analytic 
extensions for $Q^2 < Q_P^2$}
\label{appTables}

We present in Tables~\ref{tab12p1}--\ref{tab32m2}
the coefficients associated with 
all the states $\frac{1}{2}^\pm$ and  $\frac{3}{2}^\pm$ 
according to the expressions from Table~\ref{tabAmps}.
These parametrizations are used in the 
numeric results presented in Sec.~\ref{secResults}.
The numerical values of the coefficients 
are determined by the relations 
between the coefficients 
$a_l$, $b_l$ and $c_l$ and the derivatives 
in $\tilde q^2$, according to relations derived in 
Appendix~\ref{appDerivatives}.

We use $\alpha_l$ to represent $a_l$, $b_l$ and $c_l$ ($l=0,1,2,3$)
according to the corresponding amplitude.
To represent the values of $b_0$ and $c_0$ determined 
by some pseudothreshold condition we use bold.

We use also bold to represent the values of $b_3$ 
and $c_3$ when they are determined by 
continuity conditions (see Appendix~\ref{appendix-alpha}),
and not by the third derivative of the amplitudes.

The large magnitude of some coefficients 
is not a handicap because those coefficients
are multiplied by powers of $\tilde q^2$
which can be small near $Q^2=0$,
where $|{\bf q}| \simeq \frac{M_R^2-M_N^2}{2 M_R}$
and $\tilde q^2 \simeq \left( \frac{M_R^2-M_N^2}{2 M_R^2} \right)^2$.

%\clearpage

\begin{table}[b]
 %\begin{minipage}{0.5\linewidth}
  \begin{minipage}{1.0\linewidth}
  \centering
\begin{tabular}{c| r r r r }
\hline
\hline
$Q_P^2= 0.5$ & $\alpha_0$ & $\alpha_1$ & $\alpha_2$ & $\alpha_3$ \\[.1cm]
$A_{1/2}$ & $ -0.25945 $ &   0.17552  &   $- 3.95888$  &   3.61489 \\
$S_{1/2}$ & $ 0.34126$ &   $-1.1610$ &    $-1.2194$ &   0.30314  \\[.12cm]
$Q_P^2= 0.3$ & $\alpha_0$ & $\alpha_1$ & $\alpha_2$ & $\alpha_3$ \\[.1cm]
$A_{1/2}$ & $ -0.39261$ &   3.54955 &   $ - 12.0959$ &   16.0291 \\
$S_{1/2}$ & $ -0.47853$ &   $-2.3693$ &   3.8070 &   0.74123 \\[.12cm]
$Q_P^2= 0.1$ & $\alpha_0$ & $\alpha_1$ & $\alpha_2$ & $\alpha_3$ \\[.1cm]
$A_{1/2}$ & $  -0.87510$  & \spQ  13.9245  & \spQ  $ -87.7413$ & \spQ  202.738\\
$S_{1/2}$ & 0.78168 &   $-6.5333$ &   $ 18.367$ &   0.80945\\
\hline
\hline 
\end{tabular}
\centering
%\end{center}
\caption{\footnotesize
$\gamma^\ast N \to N(1440)$ amplitudes. 
$Q_P^2$ is in GeV$^2$.  \\ The  coefficients $\alpha_l$ 
are in units of GeV$^{-1/2}$.}
\label{tab12p1}
 \end{minipage}%
 %\begin{minipage}{0.5\linewidth}
 \begin{minipage}{1.0\linewidth}
  \centering
 %\begin{center}
\begin{tabular}{c| r r r r }
\hline
\hline
$Q_P^2= 0.5$ & $\alpha_0$ & $\alpha_1$ & $\alpha_2$ & $\alpha_3$ \\[.1cm]
$A_{1/2}$ &  0.14802 &  $ -0.74153$    &   1.60791 & $ -1.35352$  \\
$S_{1/2}$ & $  -0.094737$ &   $0.42481$ &    $-0.068790$ &    0.21357  \\[.12cm]
$Q_P^2= 0.3$ & $\alpha_0$ & $\alpha_1$ & $\alpha_2$ & $\alpha_3$ \\[.1cm]
%\hline
$A_{1/2}$ &  0.14591  & \spQ  $ -0.70895$  & \spQ  1.44138 & \spQ  $-1.07066$ \\
$S_{1/2}$ &  $-0.11720$ &   0.64051 &   $-1.2308$ &   0.32256 \\[.12cm]
$Q_P^2= 0.1$ & $\alpha_0$ & $\alpha_1$ & $\alpha_2$ & $\alpha_3$ \\[.1cm]
$A_{1/2}$ &   0.11288 &  $-0.08186$  &  $ -2.54812$  &  7.43066 \\
$S_{1/2}$ &  $-0.15104$ &  1.0458 &   $-2.4157$ &   0.10166 \\
\hline
\hline 
\end{tabular}
\centering
%\end{center}
\caption{\footnotesize
$\gamma^\ast N \to N(1710)$ amplitudes. 
$Q_P^2$ is in GeV$^2$. \\ The  coefficients $\alpha_l$ 
are in units of GeV$^{-1/2}$.}
\label{tab12p2}
 \end{minipage}%
\end{table}

\clearpage

\begin{table}
%\begin{center}
\begin{minipage}{1.0\linewidth}
\centering
\begin{tabular}{c| r r r r }
\hline
\hline
$Q_P^2= 0.5$ & $\alpha_0$ & $\alpha_1$ & $\alpha_2$ & $\alpha_3$ \\[.1cm]
$A_{1/2}$ &  0.090914 &   0.029760   &   $ -0.14510$  &   0.10136 \\
$S_{1/2}$ & {\bf 0.16557} &   $-2.27506$ &    9.01097 &   ${\bf -11.8996}$  \\[.12cm]
$Q_P^2= 0.3$ & $\alpha_0$ & $\alpha_1$ & $\alpha_2$ & $\alpha_3$ \\[.1cm]
%\hline
$A_{1/2}$ &   0.090772  & \spQ    0.031642 & \spQ  $- 0.15346$ & \spQ  0.11382 \\
$S_{1/2}$ &   {\bf 0.16531} &   $-3.11243$ &   1.67371 &   ${\bf -29.8249}$ \\[.12cm]
$Q_P^2= 0.1$ & $\alpha_0$ & $\alpha_1$ & $\alpha_2$ & $\alpha_3$ \\[.1cm]
$A_{1/2}$ &   0.090737  &   0.032275 &  $ -0.15734$  &  0.12185 \\
$S_{1/2}$ &   {\bf 0.16525}  &  $-4.76363$ &   38.4450 &   ${\bf -102.058}$ \\
\hline
\hline 
\end{tabular}
\centering
\caption{\footnotesize
$\gamma^\ast N \to N(1535)$ amplitudes. 
$Q_P^2$ is in GeV$^2$.  \\ The  coefficients $\alpha_l$ 
are in units of GeV$^{-1/2}$.}
\label{tab12m1}
 \end{minipage}%
%\end{table}
%\begin{table}
%\begin{center}
 \begin{minipage}{1.0\linewidth}
  \centering
\begin{tabular}{c| r r r r }
\hline
\hline
$Q_P^2= 0.5$ & $\alpha_0$ & $\alpha_1$ & $\alpha_2$ & $\alpha_3$ \\[.1cm]
$A_{1/2}$ &  ${\bf  -0.020926}$  &   1.2457   &   $-5.3063$  &   {\bf  6.8489} \\
$S_{1/2}$ &  $ -0.038110$ &   0.12146 &    $ - 0.38834$ &    0.38864 \\[.12cm]
$Q_P^2= 0.3$ & $\alpha_0$ & $\alpha_1$ & $\alpha_2$ & $\alpha_3$ \\[.1cm]
%\hline
$A_{1/2}$ &  ${\bf -0.024278 }$ & \spQ 1.8536  & \spQ  $ -9.7710$ & \spQ  {\bf  17.037} \\
$S_{1/2}$ &    $ -0.044215$ &  0.20563  &   $ -0.77796 $ &   0.99409 \\[.12cm]
$Q_P^2= 0.1$ & $\alpha_0$ & $\alpha_1$ & $\alpha_2$ & $\alpha_3$ \\[.1cm]
$A_{1/2}$ &   ${\bf -0.030073}$ &   2.8745 &  $ -22.447$  & {\bf 58.338} \\
$S_{1/2}$ &     $ -0.054768$  &   0.41255 &   $ -2.14777$ &    4.05433  \\
\hline
\hline 
\end{tabular}
\centering
%\end{center}
\caption{\footnotesize
Alternative parametrization of the 
$\gamma^\ast N \to N(1535)$ amplitudes. 
$Q_P^2$ is in GeV$^2$. The coefficients $\alpha_l$ 
are in units of GeV$^{-1/2}$.}
\label{tab12m1b}
 \end{minipage}%
\end{table}

\begin{table}
\begin{minipage}{1.0\linewidth}
\centering
%\begin{table}
%\begin{center}
\begin{tabular}{c| r r r r }
\hline
\hline
$Q_P^2= 0.5$ & $\alpha_0$ & $\alpha_1$ & $\alpha_2$ & $\alpha_3$ \\[.1cm]
$A_{1/2}$ &  0.14242 &   $-1.4491$  &    7.2241  &  $-1.0966$ \\
$S_{1/2}$ &  {\bf 0.23371} &   $-2.9066$  &    11.614 &   ${\bf -16.113}$  \\[.12cm]
$Q_P^2= 0.3$ & $\alpha_0$ & $\alpha_1$ & $\alpha_2$ & $\alpha_3$ \\[.1cm]
%\hline
$A_{1/2}$ &    0.076072 & \spQ    $-0.52719$ & \spQ  3.1199 & \spQ  $-0.47123$ \\
$S_{1/2}$ &    {\bf 0.12483} &   $-2.0636$ &   10.738 &   ${\bf -19.398}$ \\[.12cm]
$Q_P^2= 0.1$ & $\alpha_0$ & $\alpha_1$ & $\alpha_2$ & $\alpha_3$ \\[.1cm]
$A_{1/2}$ &   0.090202  &  $- 0.85008$ &   0.52608  &  $ -9.7186$ \\
$S_{1/2}$ &   {\bf 0.14802} &   $-3.3135$ &    23.664 &   ${\bf -57.280}$ \\
\hline
\hline 
\end{tabular}
\centering
%\end{center}
\caption{\footnotesize
$\gamma^\ast N \to N(1650)$ amplitudes. 
$Q_P^2$ is in GeV$^2$. \\ The  coefficients $\alpha_l$ 
are in units of GeV$^{-1/2}$.}
\label{tab12m2}
 \end{minipage}%
 \begin{minipage}{1.0 \linewidth}
  \centering
\begin{tabular}{c| r r r r }
\hline
\hline
$Q_P^2= 0.5$ & $\alpha_0$ & $\alpha_1$ & $\alpha_2$ & $\alpha_3$ \\[.1cm]
$A_{1/2}$ &  $-0.22201$ &   2.2354     &   $-3.4082$  &   633.79 \\
$S_{1/2}$ &   {\bf 0.18814}   &   $-7.9759$  &   45.401  &   ${\bf -72.894}$  \\[.12cm]
$Q_P^2= 0.3$ & $\alpha_0$ & $\alpha_1$ & $\alpha_2$ & $\alpha_3$ \\[.1cm]
%\hline
$A_{1/2}$ &    0.10377 & \spQ    $-0.60981$ & \spQ   1.2866 & \spQ  $-0.73707$ \\
$S_{1/2}$ &    {\bf 0.17455} &      $-17.903$ &   140.12 &   ${\bf -298.61}$ \\[.12cm]
$Q_P^2= 0.1$ & $\alpha_0$ & $\alpha_1$ & $\alpha_2$ & $\alpha_3$ \\[.1cm]
$A_{1/2}$ &   $-0.028634$ &   1.9674 &   $-15.565$  &  36.266 \\
$S_{1/2}$ &   ${\bf -0.048166}$ &   $ -37.796 $ &   413.95  &   ${\bf - 1186.1}$ \\
\hline
\hline 
\end{tabular}
\centering
\caption{\footnotesize
$\gamma^\ast N \to \Delta(1620)$ amplitudes. 
$Q_P^2$ is in GeV$^2$. \\ The coefficients $\alpha_l$ 
are in units of GeV$^{-1/2}$.}
\label{tab12m3}
\end{minipage}
\end{table}

\begin{table}
  \begin{minipage}{1.0\linewidth}
  \centering
%\begin{center}
\begin{tabular}{c| r r r r }
\hline
\hline
$Q_P^2= 0.5$ & $\alpha_0$ & $\alpha_1$ & $\alpha_2$ & $\alpha_3$ \\[.1cm]
$A_{1/2}$ &  $ -0.65848$  &    2.6122    &   $ -4.7555 $  &    3.3803 \\
$A_{3/2}$ &  $ -1.2512$  &     4.9645   &   $ - 9.0357$  &     6.4217 \\
$S_{1/2}$ &  {\bf 0.19008 }  &   $ -0.57303$  &   0.51477    &  {\bf 0.090141}  \\[.12cm]
$Q_P^2= 0.3$ & $\alpha_0$ & $\alpha_1$ & $\alpha_2$ & $\alpha_3$ \\[.1cm]
%\hline
$A_{1/2}$ &    $ -0.86670 $ & \spQ   0.50182  & \spQ   $ - 14.182$ & \spQ    15.893 \\
$A_{3/2}$ &    $  -1.6468$ & \spQ  9.5352 & \spQ   $ - 26.942$ & \spQ  30.192  \\
$S_{1/2}$ &   {\bf 0.24999} &   $ -0.66646$ &   $-1.5748$ &   {\bf  6.2593} \\[.12cm]
$Q_P^2= 0.1$ & $\alpha_0$ & $\alpha_1$ & $\alpha_2$ & $\alpha_3$ \\[.1cm]
$A_{1/2}$ &   $ - 1.3262$  &   15.134  &   $- 91.948$  &    223.67 \\
$A_{3/2}$ &   $  -2.5198$  &   28.755  &   $ - 174.69$  &   424.96 \\
$S_{1/2}$ &   {\bf  0.38240}   &  1.9501 &   $ -67.905$  &   {\bf 315.18} \\
\hline
\hline 
\end{tabular}\centering
%\end{center}
\caption{\footnotesize
$\gamma^\ast N \to \Delta(1232)$ amplitudes. 
$Q_P^2$ is in GeV$^2$. \\ The coefficients $\alpha_l$ 
are in units of GeV$^{-1/2}$.}
\label{tab32p1}
 \end{minipage}%
%\begin{table}
%\begin{center}
 \begin{minipage}{1.05\linewidth}
  \centering
\begin{tabular}{c| r r r r }
\hline
\hline
$Q_P^2= 0.5$ & $\alpha_0$ & $\alpha_1$ & $\alpha_2$ & $\alpha_3$ \\[.1cm]
$A_{1/2}$ &    0.79910 &     $-6.3788$   &   19.119    &     $ - 20.447$ \\
$A_{3/2}$ &   $-0.14988$   &    0.57591    &   $ -1.6760$  &    2.0676 \\
$S_{1/2}$ &  {\bf   1.3792}  &   $ -17.488$  &    73.909    &  ${\bf -109.88}$  \\[.12cm]
$Q_P^2= 0.3$ & $\alpha_0$ & $\alpha_1$ & $\alpha_2$ & $\alpha_3$ \\[.1cm]
%\hline
$A_{1/2}$ &    0.85564  & \spQ     $ -7.0744$  & \spQ   21.925 & \spQ   $ -24.134$ \\
$A_{3/2}$ &    $ -0.16596$ & \spQ    0.80116 & \spQ   $  -2.7324$ & \spQ    3.7261\\
$S_{1/2}$ &   {\bf 1.4817} &      $-23.729$ &    125.44 &   ${\bf - 225.33}$ \\[.12cm]
$Q_P^2= 0.1$ & $\alpha_0$ & $\alpha_1$ & $\alpha_2$ & $\alpha_3$ \\[.1cm]
$A_{1/2}$ &   $ -0.10077$  &    10.935 &   $ - 91.648$  &   215.69 \\
$A_{3/2}$ &   $ -0.18719$  &    1.1853 &   $  -5.0639$  &    8.4724 \\
$S_{1/2}$ &   ${\bf  0.011378}$   &  $- 0.63529$ &    4.4749 &   ${\bf - 14.022}$ \\
\hline
\hline 
\end{tabular} \centering
%\end{center}
\caption{\footnotesize
$\gamma^\ast N \to N(1720)$ amplitudes. 
$Q_P^2$ is in GeV$^2$. \\ The coefficients $\alpha_l$ 
are in units of GeV$^{-1/2}$.}
\label{tab32p2}
 \end{minipage}%
\end{table}

\clearpage

\begin{table}
%\begin{center}
 \begin{minipage}{1.0\linewidth}
  \centering
\begin{tabular}{c| r r r r }
\hline
\hline
$Q_P^2= 0.5$ & $\alpha_0$ & $\alpha_1$ & $\alpha_2$ & $\alpha_3$ \\[.1cm]
$A_{1/2}$ &    0.055412  &     $-1.0490$     &   3.0785    &      $-2.9631$ \\
$A_{3/2}$ &    {\bf 0.095976}   &    0.13925   &    $ -1.9712$    &  {\bf 3.4359} \\
$S_{1/2}$ &    ${\bf -0.20501}$ &    0.48652  &    0.54241   &  ${\bf - 2.0547}$ \\[.12cm]
$Q_P^2= 0.3$ & $\alpha_0$ & $\alpha_1$ & $\alpha_2$ & $\alpha_3$ \\[.1cm]
%\hline
$A_{1/2}$ &    0.052805 & \spQ   $-0.99861$  & \spQ  2.7810  & \spQ   $ -2.4054$ \\
$A_{3/2}$ &   {\bf  0.091461} & \spQ     0.81210 & \spQ   $ -7.9402$ & \spQ   {\bf  17.192} \\
$S_{1/2}$ &   ${\bf -0.19537}$ &   $-0.32830$    &   7.5377 &    ${\bf - 18.038}$ \\[.12cm]
$Q_P^2= 0.1$ & $\alpha_0$ & $\alpha_1$ & $\alpha_2$ & $\alpha_3$ \\[.1cm]
$A_{1/2}$ &   0.027233  &   $ -0.50818$  &   $-0.39473$  &   4.5374  \\
$A_{3/2}$ &   {\bf 0.047169}  &    0.41418   &   $ -47.084$  &   {\bf 144.70} \\
$S_{1/2}$ &   ${\bf -0.10076}$ &  $-5.4874$  &   64.794      &  ${\bf -200.36}$ \\
\hline
\hline 
\end{tabular} \centering
%\end{center}
\caption{\footnotesize
$\gamma^\ast N \to N(1520)$ amplitudes. 
$Q_P^2$ is in GeV$^2$. \\ The coefficients $\alpha_l$ 
are in units of GeV$^{-1/2}$.}
\label{tab32m1}
%\end{table}
 \end{minipage}%
%\begin{table}
%\begin{center}
 \begin{minipage}{1.0\linewidth}
  \centering
\begin{tabular}{c| r r r r }
\hline
\hline
$Q_P^2= 0.5$ & $\alpha_0$ & $\alpha_1$ & $\alpha_2$ & $\alpha_3$ \\[.1cm]
$A_{1/2}$ &    0.18100      &   \spQ     0.056698   &    \spQ  $-4.0960$      & \spQ     7.9039   \\
$A_{3/2}$ &    {\bf 0.31351}      &    $-2.5409$     &    7.2162    &  ${\bf - 6.8990}$ \\
$S_{1/2}$ &    ${\bf -0.57183}$ &     8.0388  &    $ -34.203$   &  {\bf 47.755}  \\[.12cm]
$Q_P^2= 0.3$ & $\alpha_0$ & $\alpha_1$ & $\alpha_2$ & $\alpha_3$ \\[.1cm]
%\hline
$A_{1/2}$ &    $-0.10410$ &      3.9660 &   $-22.030$  &    35.433 \\
$A_{3/2}$ &   ${\bf  -0.18031}$ &     5.9459  &    $-40.325$ &    {\bf  80.479} \\
$S_{1/2}$ &   ${\bf 0.32887}$ &   $ -3.8547$    &   18.709 &    ${\bf - 31.637}$ \\[.12cm]
$Q_P^2= 0.1$ & $\alpha_0$ & $\alpha_1$ & $\alpha_2$ & $\alpha_3$ \\[.1cm]
$A_{1/2}$ &    0.0527443 &    3.90692  &   $- 47.7833$  &    $-327.954$ \\
$A_{3/2}$ &   {\bf 0.091356}  &    0.66483    &   $ -3.0664$  &   ${\bf -12.083}$ \\
$S_{1/2}$ &   ${\bf - 0.16663}$ &   5.4626 &       $-40.781$   &   {\bf 97.083} \\
\hline
\hline 
\end{tabular} \centering
%\end{center}
\caption{\footnotesize
$\gamma^\ast N \to \Delta(1700)$ amplitudes. 
$Q_P^2$ is in GeV$^2$. \\ The coefficients $\alpha_l$ 
are in units of GeV$^{-1/2}$.}
\label{tab32m2}
 \end{minipage}%
\end{table}

%\clearpage
%\input{biblo}

\end{document}